\documentclass[draft]{agujournal2025}

\usepackage{url} 
\usepackage{lineno}
\usepackage[inline]{trackchanges} 
\usepackage{soul}
\usepackage{enumitem}
\usepackage{multirow}
\usepackage{makecell}
\usepackage{longtable}

\usepackage{xspace}
\usepackage{amsmath}

\newcommand*{\ie}{i.e.\@\xspace}

\usepackage{todonotes}

\draftfalse

\makeatletter

\def\thejournalname{}

\def\ps@headings{%
  \def\@oddfoot{\centerline{\small --\the\c@page--}}%
  \let\@evenfoot\@oddfoot
  \def\@oddhead{}%
  \let\@evenhead\@oddhead
}
\ps@headings

\gdef\typesetjournal{}

\makeatother

\nolinenumbers
\begin{document}

\title{How well is surface ocean carbon represented in observations and ocean models?}

\authors{%
                  Viviana Acquaviva\affil{1,4},
                  Romina Wild\affil{2}, Alessandro Laio\affil{3}, Amanda R. Fay\affil{4}, Thea H. Heimdal\affil{4}, Galen A. McKinley\affil{4} \
	      }

\affiliation{1}{Physics Department, New York City College of Technology, 300 Jay Street, Brooklyn, 11201, NY, USA} 

\affiliation{2}{National Institute of Oceanography and Applied Geophysics
 (OGS), Borgo Grotta Gigante 42/C, Sgonico (TS), 34010, Italy}

\affiliation{3}{Scuola Internazionale Superiore di Studi Avanzati (SISSA), Via Bonomea, 265, Trieste, 34136,  Italy}

\affiliation{4}{Columbia University and Lamont-Doherty Earth Observatory, 61 Route 9W, Palisades, 10964, New York, USA}

\correspondingauthor{Viviana Acquaviva}{vacquaviva@citytech.cuny.edu}



\section*{Abstract}

We introduce a general framework for quantifying the information content and representation quality of complex geophysical datasets based on the intrinsic dimension and differentiable information imbalance of data manifolds. We use it to derive and compare optimal representations of surface ocean carbon in the SOCAT database of observations and in global ocean biogeochemistry models (GOBMs) and to assess the robustness of the information we can extract from existing data. We find that within the most widely used feature set, the complexity of the data space of SOCAT observations is not fully captured by GOBMs, but the ranking and relative importance of variables learned through GOBMs are substantially correct. We observe that the learned representation of ocean carbon is less accurate in some regions, including the Southern Ocean, but doesn't appear to have evolved significantly over the last two decades. Finally, we show how the optimal representations can be used to improve the skill of distance-based machine learning models and demonstrate it for ocean carbon, and we propose two new metrics to compare models and observations that can be used to build more accurate weighted ensembles of estimates.






\section{Introduction}

The ocean plays a major role in the global carbon cycle by removing carbon from the atmosphere, absorbing about a quarter of global human carbon emissions \cite{DeVries2022}. However, large uncertainties remain.
The ocean sink is often estimated through two main methods: 1) observation-based data products, or 2) global ocean biogeochemical models (GOBMs). Data products reconstruct the surface ocean CO$_2$ concentration (often expressed in terms of its fugacity, fCO$_2$) from sparse observations combined with full-coverage satellite and reanalysis data through a machine learning (ML) approach.  The reconstructed full CO$_2$ fields are then used with reanalysis products of wind speed and other physical variables to obtain air-sea carbon fluxes. The average global ocean cxarbon sink as estimated from these two sources and, with some adjustments, is reported by the annual Global Carbon Budget (GCB; \citeA{Friedlingstein2025, Friedlingstein2026nature}). The spread among the ensembles of GOBMs estimates and data products is added in quadrature and used as an estimate of flux uncertainty \cite{Heimdal2025}. More accurate estimates of the ocean carbon sink directly translate into reduced uncertainties in the global carbon cycle, and in turn, into more efficient planning for mitigation of and adaptation to climate risks \cite{Peters2017, McKinley2026}.

In this paper, we apply two novel statistical tools, the intrinsic dimension (ID, \citeA{Denti2022}) and the differentiable information imbalance (DII, \citeA{Wild2025}), to learn the representation of CO$_2$ in observations from the SOCAT database \cite{SOCATdataset, SOCAT} 
and compare it to the one derived from GOBMs. The ID aims to assess the complexity of data spaces, while the DII is used to compare the metric spaces of input features (spatio-temporal variables and satellite-based observations) and target variables (CO$_2$). These tools, applied to ocean carbon for the first time, allow us to find optimal representations, to assess the residual expected error in modeling a target as a function of the input features (in other words, the information imbalance between the two spaces), and to define new metrics for model-observation comparison or intermodel comparison. We note that while our results can be directly employed to improve machine learning methods that use distances (see Sec. \ref{sec:ML}), they are only based on the information content of different data spaces and do not employ machine learning modeling.

To apply our framework, we select the input variables commonly used to build machine learning models of the CO$_2$ field (salinity, sea surface temperature, Chlorophyll concentration, mixed layer depth, atmospheric concentration of CO$_2$, and several spatio-temporal variables; see Sec. \ref{sec:setup} for details). We define as target variable $\Delta$CO$_2$, {\it i.e.} the difference between the fugacity of CO$_2$ and the concentration of atmospheric CO$_2$, following the approach of \citeA{Gregor2024,Fay2024,heimdal2026update}. Then, we develop statistically meaningful metrics to compare how $\Delta$CO$_2$ is represented in models and in direct observations.

We consider several use cases. First, we compare observations from SOCAT to five different GOBMs in the same spatial and temporal domain. We use two primary statistical tools to compare the complexity of the data manifolds in models and observations, the skill of input variables in predicting $\Delta$CO$_2$, and the ranking of relevant variables. We propose new metrics to compare the representation of $\Delta$CO$_2$ in the GOBMs to that in the observations, and derive conclusions on which models are closest to observations.

Next, we assess the effect of the sparsity and bias of the existing SOCAT observations, which are not uniformly distributed in space and time, on the learned representation of $\Delta$CO$_2$, asking whether it is robust to generalization outside the learning domain. In GOBMs, this can be done straightforwardly by assessing how well the representation learned within the SOCAT domain translates to a global, uniform spatio-temporal domain. For observations, where no global uniform sample is present, we run experiments where we learn the representation of $\Delta$CO$_2$ in one domain and apply to a different subdomain, defined either through ablation (removing observations in certain areas) or weighted resampling. We also derive a general comparison framework to assess inter-model differences for GOBMs. 

Last, we show how our framework can be used to improve the performance of distance-based machine learning models of $\Delta$CO$_2$ (such as clustering methods or neural networks), by building a $k$-Nearest Neighbor model based on our optimized distance metric, and comparing it to the one based on Euclidean distance in feature space.

The paper is organized as follows. In Sec. \ref{sec:GOBMs}, we describe the GOBMs that we use and how they participate in the Global Carbon Budget. In Sec. \ref{sec:tools}, we present the statistical tools that we use to study the feature space and its relation to the target space, most notably the intrinsic dimension estimate \cite{Denti2022} and the differential information imbalance \cite{Wild2025}. After presenting our analysis set up in Sec. \ref{sec:setup}, we analyze SOCAT observations to derive an optimal feature set in Sec. \ref{sec:SOCAT_only}, and compare observations and GOBMs and assess model performance in \ref{sec:models}. We present the effect of data sparsity and bias, both in observations and GOBMs, in Sec. \ref{sec:sparsity}. We show our application to machine learning modeling in Sec. \ref{sec:ML}. 

\section{SOCAT and GOBMs in GCB}
\label{sec:GOBMs}
\subsection{GOBMs}
GOBMs provide a process-based representation of the marine carbon cycle and are a key component of the GCB framework to estimate ocean carbon uptake. GOBMs simulate the coupled physical, chemical and biological processes that govern the exchange and storage of carbon in the ocean, including natural and anthropogenic carbon fluxes. Within the GCB, the ensemble average of approximately 10 GOBMs is used as one of two inputs to the ocean carbon sink estimate. Models explicitly represent the transport of carbon from the surface ocean to the ocean interior, a fundamental process controlling long-term ocean carbon uptake \cite{Gruber2023}. 

This study uses monthly $1^{\circ} \times 1^{\circ}$ output of five GOBMs included in the Global Carbon Budget framework: CESM-ETHZ \cite{YangGruber2016}, FESOM-REcoM \cite{Hauck2020}, NorESM \cite{Schwinger2016}, MRI \cite{Urakawa2020}, and IPSL \cite{Aumont2015}. These models were selected because the required feature-variable output files were available for all models, enabling a consistent comparative machine-learning analysis across multiple modeling frameworks \cite{Roobaert2026preprint}.

\subsection{SOCAT}
Surface ocean CO$_2$ observations are obtained from the Surface Ocean CO$_2$ Atlas (SOCAT) version 2025 \cite{SOCATdataset, SOCAT}, a community-compiled database of quality-controlled in situ surface ocean fugacity of CO$_2$ (fCO$_2$) measurements collected from ships, moorings, and autonomous observing platforms worldwide. In this study, we used the monthly gridded product, which aggregates individual observations onto a common 1° × 1° spatial grid and monthly temporal resolution.

In the SOCAT dataset, the distribution of measurements is sparse and heterogeneous in both space and time. At 1° × 1° monthly resolution, direct observations occupy only a small fraction (a few \%) of the global ocean, leaving substantial coverage gaps. This sparse sampling motivates the use of interpolation and machine-learning approaches to reconstruct global surface ocean carbon fields from the limited observational record \cite{Roobaert2026preprint,Hauck2020,McKinley2026local}. In this study, we step back from the machine learning reconstruction algorithms to investigate the statistical robustness of the feature setup used in these approaches.

\section{Tools}

\label{sec:tools}

\begin{figure}[h]
    \centering
    \includegraphics[width=\linewidth]{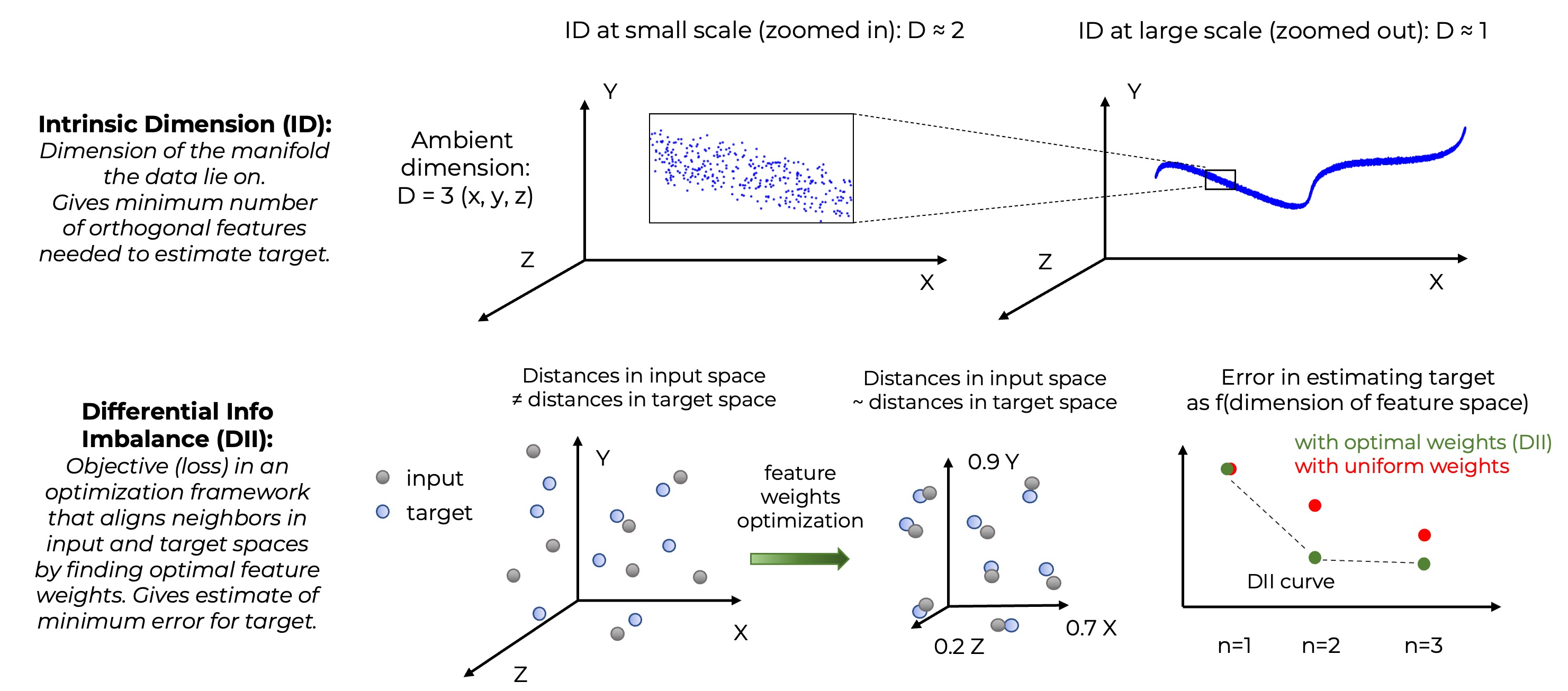}
    \caption{\textit{Top panel}: Data points are embedded in a three-dimensional space, but the dimension of the manifold where they live, \ie the ID, is lower and scale-dependent. At small scales, it is $\sim$ 2 (box on the left, inset); at large scales, it is $\sim$ 1 (box on the right). \textit{Bottom panel}: In Euclidean space, neighboring points in input space and in target space are not aligned (left panel). The DII formalism allows one to find optimal weights for the features so that neighbors are aligned in the two spaces (middle panel). The DII can be estimated for any dimension (DII curve, right panel) and quantifies the minimum information loss between input and target spaces. Low-weight features (Z, middle panel) induce a plateau in the DII curve (seen between n = 2 and n = 3 in right panel) and can be safely discarded through feature selection. A broader glossary is included in Table \ref{supp:tab:notation}.}
    \label{fig:IDinfographics}
\end{figure}

In many real-world datasets, the number of observed variables (often called ``features" in the context of ML modeling) is much larger than the number of independent directions along which the data actually vary. For example, in a dataset of handwritten digits (0–9) where each image is made of $10 \times 10$ pixels, the data space formally has a dimensionality of 100 (as each pixel is a feature), but the information needed to describe and identify the digits only varies along $<$ 10 independent directions, capturing digit identity, slant, stroke thickness, and similar factors. The intrinsic dimension of the dataset is therefore far lower than the number of pixel features \cite{Denti2022}. This effective number of degrees of freedom is known as the {\bf intrinsic dimension (ID)} of the dataset. In essence, the ID reflects the dimensionality of the manifold on which the data approximately lie, even though they are embedded in a higher dimensional space \cite{Facco2017}, as shown in Fig. \ref{fig:IDinfographics}. Estimating the ID is a valuable tool for understanding the complexity and information content of the data, as it sets a lower bound on the number of independent variables needed to represent it. For example, if the intrinsic dimension is estimated to be 5, at least 5 
uncorrelated variables are required to adequately describe the data structure. 
By comparing the ID of the data with the number of physical variables needed to build a good predictive model, we can also draw some conclusions about the efficiency of our representation; the smaller the difference between the identified ID and available dimensions of physical space, the more effective the representation is. 

The intrinsic dimension is typically scale-dependent, where scale can be thought of as the ``zoom-in" level. For example, one can have data points which lie on a line, whose coordinates are perturbed by a two-dimensional small-scale noise. In this case, the ID at small scale (zooming in) will be of order two, but at large scale (zooming out) will be of order one. This example is illustrated in the top panel of Fig. \ref{fig:IDinfographics}.

When plotting ID as a function of the scale, a plateau might appear once the true dimensionality of the local data manifold is estimated. In certain cases, no plateau might appear, for example, if the data are not contained in a well-defined low-dimensional manifold.

The ID can be estimated using different approaches. In this paper, we use the Generalized Ratios Intrinsic Dimension Estimator or GRIDE \cite{Denti2022}. This method 
uses higher-order neighbor distances to probe different scales directly on the full dataset. Specifically, it 
compares the $n$-th and $2n$-th nearest neighbor distances.
By increasing $n$, the GRIDE estimator evaluates the ID at progressively larger scales and reveals the presence (or absence) of a plateau.

In practical applications, we can estimate the intrinsic dimension 
for different configurations of input features, and evaluate how much independent information is added by each. If the estimated ID including the new feature is one full ID unit higher, then that feature carries completely independent information not represented by the other variables. If the ID remains the same, the feature adds no independent information and the manifold of other variables contains the new feature.

In this work, we use the ID in several ways. One is to compare the IDs of the space of input variables versus the space of the input variables plus the target ($\Delta$CO$_2$), in order to quantify how much independent information is carried by the target. We also use the ID to assess the similarity between observations and GOBMs, by comparing the IDs of observations and model spaces. Finally, we use it as an indicator of statistical difference between the SOCAT domain and the ``full" (uniform) domain in the GOBMs.

The second tool we use in this work is the {\bf Differentiable Information Imbalance (DII)} \cite{Wild2025}, which extends the Information Imbalance (II)) framework introduced in \citeA{Glielmo2022}. 
 These are automated methods for quantifying how informative a set of input features is with respect to a  target variable (here $\Delta$CO$_2$) in a fully interpretable manner.

Both metrics are based on the idea of neighbor ranking. We consider metric spaces A and B; in our case the space of input features (A) and target variable (B). Each data point in a data set has an index $i$ that identifies it both in space A and space B. In space A, we can rank all other data points according to how close they are to point $i$. The closest data point, let's say of index $j$, will have rank 1 (i.e., r$^A_{ij}$ = 1); the second closest will have rank 2, and so on. If metric space A is informative of metric space B (i.e., if input features contain information about the target variable), the ranking in the two spaces will align, and close neighbors in feature space will also be close neighbors in $\Delta$CO$_2$ space, making the input feature space A a good proxy for target space B. 

The II quantifies this relationship and can be approximated, for a data set with N points, as \textbf{the average neighbor rank in space B of the closest neighbor in space A:}

\begin{equation}
    II
(A \rightarrow B) \approx \frac{2\,\langle r^{B} \mid r^{A}=1\rangle}{N}.
\end{equation}

If, on average, data points that are closest in the input feature space ($r^{A}$=1) are also close in the target space, $r^{B}$ will be a small integer, and the value of the information imbalance will be close to zero. This indicates that the input features contain valuable information about the target $\Delta$CO$_2$. Conversely, if the input features contain no relevant information about the target variable, the distribution of neighbor ranks $r^B$ will be approximately uniform with a mean of $N/2$, and the information imbalance will approach 1. 

We can also interpret the numerical value of the imbalance more quantitatively: a value of 0.2 indicates that, on average, the average neighbor rank $r^{B}$ for points whose rank $r^{A} = 1$  is N/10, \ie points that are closest neighbors in the input feature space are found among the first 10\% of neighbors in the target space. As an empirical rule, for a value $x$ of II, closest neighbors in input space are found in the first $(x^{th}/2)$ percentile of neighbors in target space.

In this paper, we use a more general version of the II known as the Differentiable Information Imbalance (DII), defined as:

\begin{equation}
    DII\left(d^A\rightarrow d^B\right) = \frac{2}{N^2}\,\sum_{\substack{i,j=1 (j\neq i)}}^N c_{ij}(\lambda,d^A) \,r_{ij}^B,
\label{eq:DII}
\end{equation}

where the calculation is extended to $n$ nearest neighbors instead of only the closest one, and the softmax coefficients $c_{ij}$ assign exponentially decaying weights to neighbors based on their rank in $d^A$, modulated by the parameter $\lambda$ that describes how many nearest neighbors are included. 

The DII can be used as the objective function in an optimization framework to identify the best possible set of weighted features to predict a given target. For this purpose, we sample the space of possible weights, which are scaling factors for each feature, and find optimal weights by using a gradient-based algorithm that minimizes the DII score. The result of this operation is shown in the bottom row of Fig. \ref{fig:IDinfographics}. Optimal weights induce maximal alignment between neighbor ranks in feature space and neighbor ranks in target space; equivalently, finding optimal weights correspond to finding the most informative input metric space to predict a target.
By iterating this process for varying dimensionality of the input space, we can identify an optimal representation and assess how the information content of the representation varies as features are progressively eliminated. Thus, the differentiable objective introduced by the DII enables a pipeline for ranking the importance of the features in high-dimensional systems, as well as for feature selection.

In this work, we use the DII to identify feature sets which are maximally informative toward the $\Delta$CO$_2$ target and to estimate how those feature sets vary across observations, models and spatiotemporal domains. A glossary of the terms presented in this section is presented in Table \ref{supp:tab:notation}. We use the implementation of the ID, II, and DII available in the \textit{dadapy} package \cite{dadapy}.

\section{Features and target setup}
\label{sec:setup}

We consider an input space of 13 variables, chosen to match those used in some methods to reconstruct global fields of fCO$_2$ using machine learning, for example in \cite{bennington2022, Gloege2022, Chau2022, Landschutzer_2016}. Four of these variables describe the physical and biogeochemical state of the surface ocean: sea surface temperature, sea surface salinity, mixed layer depth, and chlorophyll-a concentration in logarithmic units. 
Additionally, three variables are defined as anomalies from the monthly climatology, \ie the monthly average of each variable over the time span of 1982-2024. 
The only input variable representing an atmospheric field is the time-varying global monthly atmospheric concentration of carbon dioxide, which is obtained from the NOAA Global Monitoring Laboratory. Finally, three variables describe the location on the Earth sphere (A, B, C), in units engineered to have periodic boundary conditions (in other words, to encode the fact that longitude 0$^\circ$ and 360$^\circ$ are the same), and two describe the time of the year (T$_0$ and T$_1$), again using an embedding chosen to enforce periodic conditions. For all variables, we use a gridded 1$^\circ$ by 1$^\circ$ product with monthly records. The full description of the variables is provided in Table \ref{supp:tab:variables}.

As target variable, we choose $\Delta$CO$_2$, the difference between the fugacity of CO$_2$ in the surface ocean and the concentration of CO$_2$ in the atmosphere: $\Delta$CO$_2 \, = \, $fCO$_2 \, - \, $xCO$_2$. In this definition, xCO$_2$  varies with time, but has been averaged across the spatial dimensions. The use of $\Delta$CO$_2$ as a target aims to remove the long-term trend of ocean fCO$_2$, alleviating the issue of modeling an evolving target; further work by our group \cite{heimdal2026update} and others \cite{Gregor2024,Wanninkhof2025} look at the benefits of this choice in more detail.

Minimizing the differentiable information imbalance for very large data sets poses some computational challenges. For a data set of size $N$, we need to calculate and store a $N \times N$ matrix of distances for all possible combinations of features and weights. This requires both a large RAM allocation and a large amount of CPU time. After extensive testing, we chose our primary study period of three years, from January 2020 to December 2022, and considered five random subsamples of 16,000 points for observations and model outputs. In section \ref{sec:time_biomes}, we evaluate the sensitivity of our results to this choice of timeframe.  We tested that the intrinsic dimensions, feature sets, and weights selected by the differential information imbalance have converged for this sample set size to within 5\% for all case studies presented here (see Sec. \ref{sec:Convergence}). Data sets are standardized by removing the mean and dividing by the standard deviation of each feature before processing. Even though the DII can also handle non-standardized data and find the optimal scaling between features, normalizing the data first largely removes the component of the weight due to unit correction, and thus the weights can be interpreted as relative importance scaling between features.

\section{Optimal metric space to predict $\Delta$CO$_2$ in SOCAT observations}
\label{sec:SOCAT_only}

In this first section, we consider observations from the SOCAT database \cite{SOCATdataset, SOCAT}. The aim is to determine how well different sets of input features describe observed $\Delta$CO$_2$, how many features are necessary for this task, and what their relative importance is (feature weights). Therefore, we examine the \textit{intrinsic dimensionality} (ID) of the feature space and then derive feature importance weights using the \textit{differentiable information imbalance} (DII) method, introduced in Sec. \ref{sec:tools}.

\begin{figure}[ht!]
    \centering
    \includegraphics[width=\linewidth]{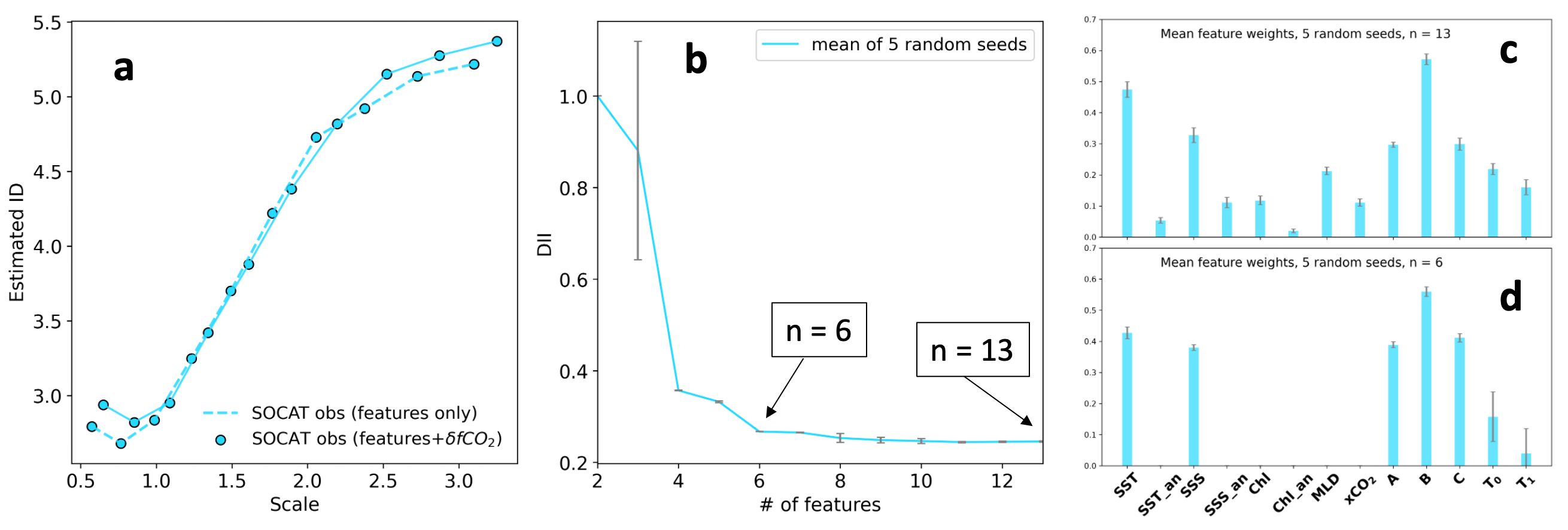}
    \caption{Analysis of SOCAT observations and input features. \textit{a)}: Intrinsic dimensionality (ID) of the SOCAT observations space. The dashed line refers to the space of the thirteen input features; the solid lines refer to the space with the addition of the target variable, $\Delta$CO$_2$. The small gap between solid and dashed lines indicates that the input variables are able to reasonably predict the target. \textit{b):} The DII measures the imbalance of information between the spaces of input features and target.  The y-axis encodes information loss as DII, and the x-axis shows input feature tuples of increasing size. There is a plateau at $n$ = 6, indicating that six variables mostly contribute to information. The residual DII of $\sim$ 0.25 indicates that some information about $\Delta$CO$_2$ is not captured by the input features. \textit{c)} and \textit{d):} Importance weights for input features towards the representation $\Delta$CO$_2$, as derived from DII optimization, with (c) the relative importance if all 13 features are considered, and (d)  the relative importance of features when only the most important 6 features are considered.} 
    \label{fig:SOCAT_only}
\end{figure}

In panel a of Fig. \ref{fig:SOCAT_only}, we show as dashed lines the IDs of the input space (the thirteen variables) and as solid lines the input+target space, which includes the target variable $\Delta$CO$_2$. 
The ID of the full (input+target) space changes from $\sim 3$ at small scale to around 5.5 at large scale. This means that, while at small scale correlations are strong, at intermediate/large scale the information content of the space of all SOCAT features could be described by between five and six independent variables. Those, of course, would not necessarily correspond to five specific physical features. However, this result suggests the presence of redundancy in the input space.

In the same figure we also report the ID for a data space that only includes the input variables but not the target variables (dashed line).  A significant gap between the ID estimated for the input variables alone and the input variable together with the target variable would indicate that the target variable is  orthogonal to the input variables, so its addition increases the ID significantly.  A small  gap, on the other hand, suggests that a large amount of information is shared between the input and target variables. In our case, the gap is approximately 0.1 in ID units (panel a of Fig. \ref{fig:SOCAT_only}). This means that including the target feature, $\Delta$CO$_2$, does not significantly increase the ID of the data manifold or, equivalently, there exists a (possibly nonlinear) function of the input features which describes $\Delta$CO$_2$ with a small error. 

Next, we seek the subset of input variables that is best suited to describe the data manifold and convey information about $\Delta$CO$_2$. In panel b of Fig. \ref{fig:SOCAT_only}, we show the DII  between the input space and $\Delta$CO$_2$ for the SOCAT observations. As specified in Sec. \ref{sec:tools}, the DII is calculated by building an optimal metric space for all possible dimensions of input variables, beginning with the full space ($n$ = 13) and progressively eliminating the least important variable. In the following, we sometimes refer to the DII for the full-dimensional space as the ``residual" DII, \ie the irreducible information imbalance between input and target data spaces. Uncertainties in the DII curve and the weights (Fig. \ref{fig:SOCAT_only}, panels c and d) are obtained by repeating the process with 5 different random seeds and reporting the mean and standard deviation of the results. 

Remarkably, the DII does not decrease significantly if one includes more than 6 variables in the input space: it is $\sim$0.24 for the full space and $\sim$0.27 for n = 6. As explained in Sec. \ref{sec:tools}, this result indicates that on average, for any data point, its closest neighbor in the input space is found within the first 12-14th percentiles of neighbors in the target CO$_2$ space. The relatively small value of the DII indicates that the scaled 6-feature input space is a good proxy for the target. We refer to n = 6 as the \textit{smallest fully informative data space}: we can safely disregard seven variables without compromising the information content (i.e., without significantly increasing the DII). An additional advantage of lower-dimensional data spaces is that their features will be less correlated than the full space, where there is a higher chance for redundancy or shared meaning. Therefore, we usually refer to the smallest fully informative space (here n = 6) when we comment on feature importances and optimal representations.

Finally, in Fig. \ref{fig:SOCAT_only} c and d, we show the relative importance of features for describing $\Delta$CO$_2$ in SOCAT observations, where relative importance is quantified by the feature weights derived from DII optimization. Panel c shows feature importance when all 13 input features are considered, panel d when enforcing sparsity with only 6 features. In the full 13-feature space, spatial feature B and the sea surface temperature (SST) emerge as particularly important. This remains true for the optimal 6-plet, where other relative importances change slightly. We also note that in panel d, seven features are shown, and the importance of features T$_0$ and T$_1$ show large error bars. This happens because in the five random samples, sometimes the optimal set includes T$_0$, and sometimes it includes T$_1$. 

\section{Comparison of $\Delta$CO$_2$ modeling in GOBMs and SOCAT observations}
\label{sec:models}

In this section, we compare the feature-target relationships observed in SOCAT observations with those predicted in five different Global Ocean Biogeochemical Models (GOBMs). Like with SOCAT observational data, we want to see how well input features derived from the GOBMs describe the resulting $\Delta$CO$_2$, using the ID and the DII methods. This allows us to assess the consistency of the models with real observations. To allow a direct comparison, the analysis in this Section is restricted to the SOCAT domain: even though all the models we considered have output at all points in space and time, we consider the samples which have the same spatial and temporal distribution of real observations, and again choose five random seeds to quantify the uncertainties due to sample variance. We will compare the SOCAT domain to the full domain of GOBMs in Sec. \ref{sec:sparsity}. 

\subsection{Intrinsic dimension of observations and GOBM spaces}

\begin{figure}
    \centering
    \includegraphics[width=\linewidth]{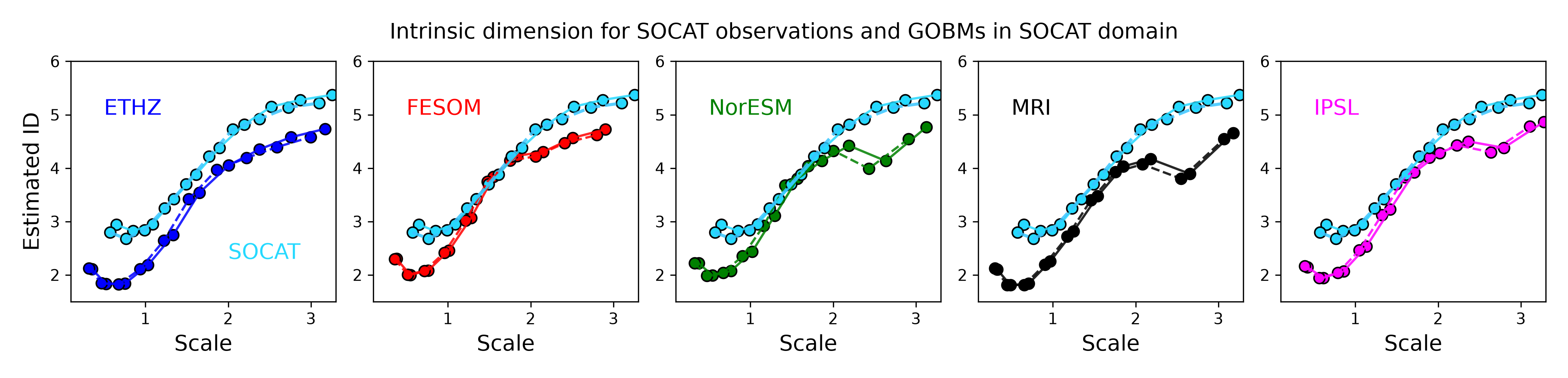}
    \includegraphics[width=\linewidth]{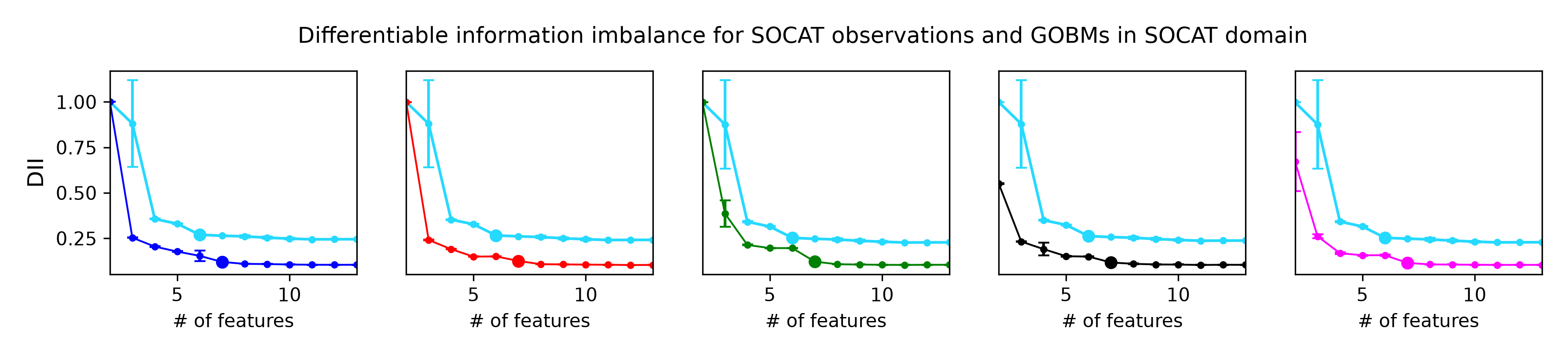}
    \caption{Top: Intrinsic dimension of the SOCAT observations space (light blue, same in all panels) and the five GOBMS we consider, limited to the same spatio-temporal domain. The dashed lines refer to the space of the thirteen input features; the solid lines refer to the space with the addition of the target variable, $\Delta$CO$_2$. The ID of the observation space is reported in all panels, to facilitate the comparison. Bottom: DII derived from SOCAT observations (light blue lines; same in each panel) and from each model. The DII is significantly larger in the observations than in models, indicating that input variables are less predictive in the real world than in the GOBMs, independently of the size of input space. The plateau at $n = 6$ for SOCAT observations and at $n = 7$ for GOBMs is marked with a larger marker. 
   }
    \label{fig:ID_Socat_models}
\end{figure}

In Fig. \ref{fig:ID_Socat_models} (top row), we plot the ID as a function of scale for each GOBM. Dashed lines represent the IDs of the input space only, and solid lines the input+target space, which includes the target variable $\Delta$CO$_2$. An important commonality is that the ID of the full (input+target) space is significantly lower, around 4.5 for all models, than the the ID estimated for real observations (light blue). This indicates that the data space of observations has (approximately) one extra degree of freedom with respect to the data spaces generated by the  models. This is an important result, which we return to in our conclusions.

The gap between the ID of the input and input+target spaces is very small ($< 0.1$ for all GOBMs,  which means that the input variables are meaningful predictors of the target in all models. This result is akin to our findings for SOCAT observations (Fig. \ref{fig:SOCAT_only}); we analyze it more quantitatively in the next subsection. 

\subsection{DII and feature weights in observations and GOBMs}
\label{sec:DII_SOCAT_models}

In Fig. \ref{fig:ID_Socat_models} (bottom row), we plot the DII as a function of the number of variables for each GOBM and compare it to the DII of SOCAT observations. An encouraging sign of consistency is the plateau at $n \, \sim$ 7, not far from the plateau at $n \, \sim$ 6 found in observations. 

However, a notable difference emerges: the DII curve is consistently lower in models than in observations. The DII for SOCAT observations was $\sim$ 0.24 - 0.27 for $n > 6$; conversely, the DII for GOBMs is $0.1 – 0.12$ for
$n>7$, indicating that the first neighbors in feature space are typically found within the first $5–6\%$ of neighbors in the target space. This means that input variables are substantially better predictors of $\Delta$CO$_2$ in GOBMs than in observations. In other words, observations of $\Delta$CO$_2$ exhibit significant more complexity that the models cannot capture. This is consistent with the smaller value of the ID observed in all the models as well. 

\begin{figure}[h!]
    \centering
    \includegraphics[width = \linewidth]{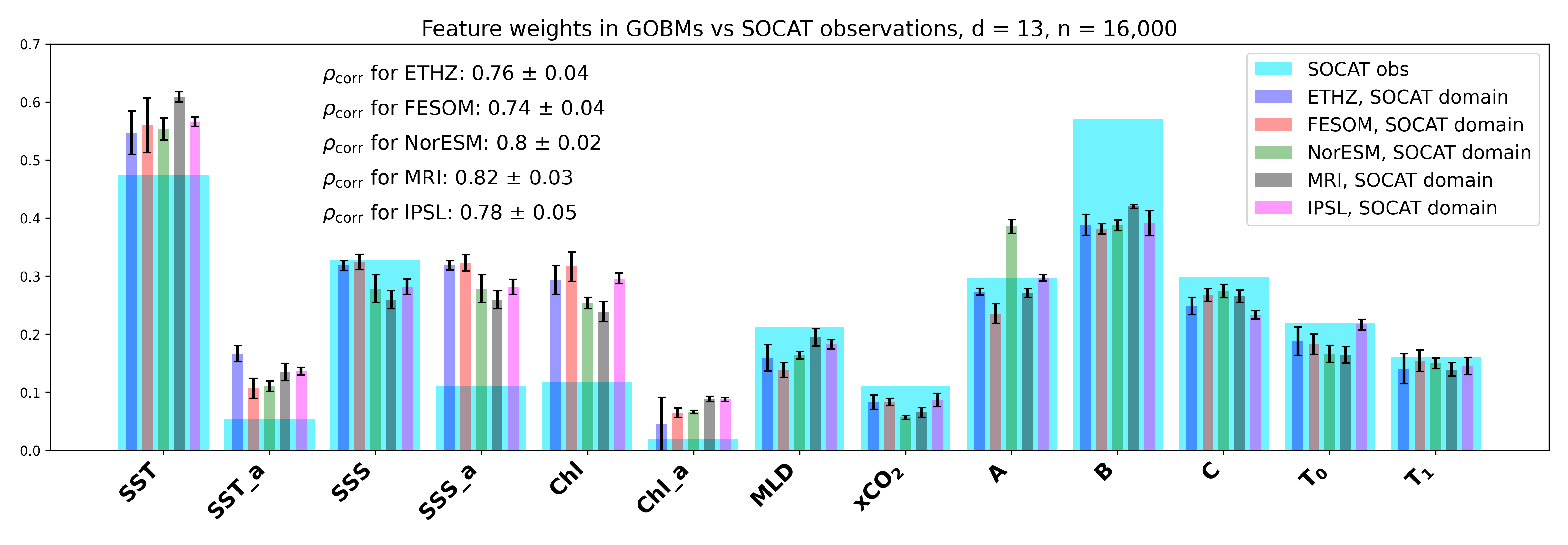}
    \includegraphics[width = \linewidth]{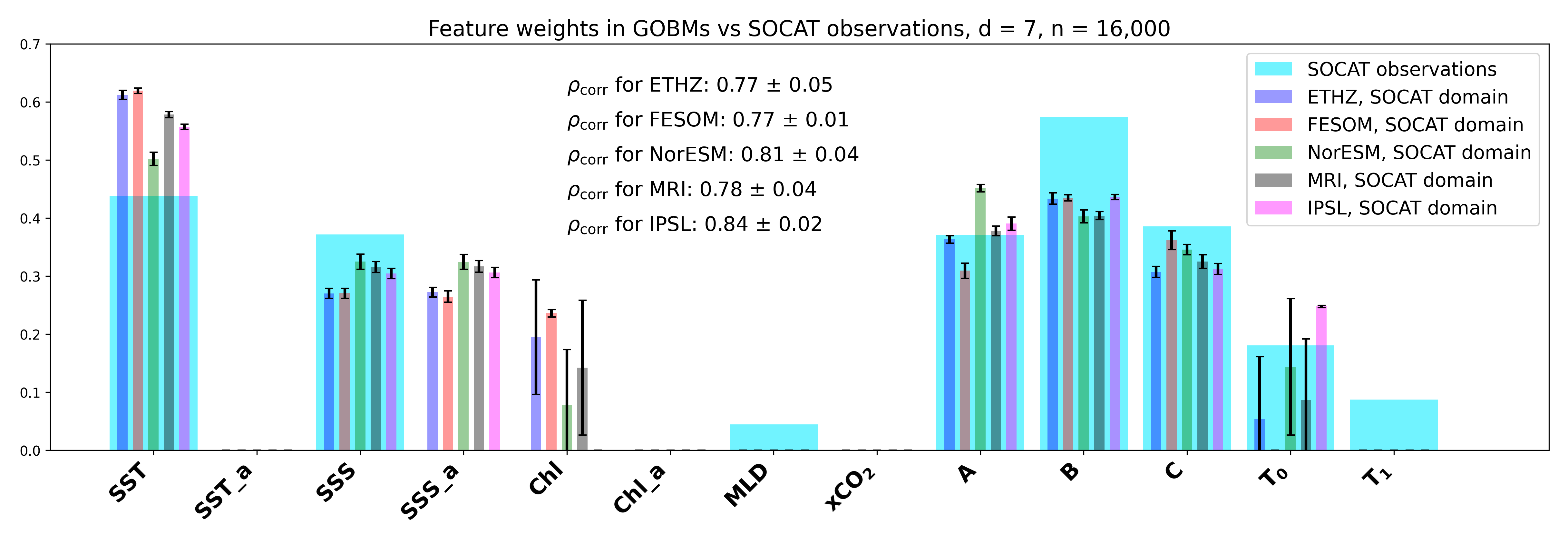}
    \caption{Feature weights for the SOCAT data (light blue) and the five GOBMs for the full (n = 13; top) dimensional space and the smallest highly informative space of observations (n = 7; bottom). For both observations and models, we show the mean over 5 random data set realizations; in models we also show the standard deviation with black error bars.  Models overestimate the importance of SSS-anomaly, SST, and Chl, while not fully capturing the importance of spatial variable B observed in SOCAT observations. The correlation coefficients between weights vectors quantify the differences. We argue that the comparison in the $n$ = 7 space is more meaningful.}
    \label{fig:SOCATweights}
\end{figure}

To gain further insights into the model-observation comparison, in Fig. \ref{fig:SOCATweights} we compare the weights obtained by optimizing the DII in the GOBMs and in SOCAT observations. Those weights provide a measure of the importance of each feature to describe $\Delta$CO$_2$. The top panel shows the weights and their errorbars (the standard deviation of weights across the five random seed)  when all 13 input features are considered, while the bottom panel shows the weights when we use only the 7 most important features. We quantify the agreement between feature weights using the Pearson correlation coefficient calculated on weights vectors. When considering either the full 13-dimensional space (top row) or the smallest fully information 7-dimensional space (bottom row), the feature weights for the five models are largely in agreement between each other. The agreement with observations, on the other hand, is good for several variables (for example, salinity, mixed layer depth, A, C), but poorer for others, with overall correlations around $\sim 0.75-0.8$ for all models. Focusing on the $n = 7$ space for its lower redundancy, we note that in SOCAT observations, spatial feature B, which combines latitude and longitude, is systematically more important than in GOBMs, while sea surface temperature (SST), salinity anomaly (SSS$_{an}$), and Chlorophyll (Chl) are less important. The optimal 7-plets for describing $\Delta$CO$_2$ vary slightly between models, but this variation should not be over-interpreted, as it can be at least partially attributed to fluctuation across different random seeds, shown by the error bars in Fig. \ref{fig:SOCATweights}. As observed in the previous section, for different random seeds, the features in the optimal 7-plet may vary, which explains why eight features are shown in the bottom panel of Fig. \ref{fig:SOCATweights}.

Overall, the inter-model agreement is higher than the agreement between models and observations, possibly indicating a common deficiency in all the GOBMs we consider. We report the ranking of features progressively eliminated during the optimal DII search as a function of $n$, for each GOBM and the SOCAT observations in Table \ref{supp:tab:SOCATrankings}. This provides a framework to compare representations across GOBMs and with observations for any dimensionality of the input space.

Finally, we propose a new metric to compare representations in different spaces, i.e., the correlation coefficient between weights derived in the two spaces for the smallest fully informative dimension (here $n$ = 7). As we mentioned in Sec. \ref{sec:SOCAT_only}, by choosing the smallest instead of the full space, we reduce (albeit not eliminate) the redundancy between different equivalent representations introduced by correlation among variables. We use this metric to compare GOBMs with SOCAT observations, concluding that in the 2020-2022 time frame, the representation of $\Delta$CO$_2$ in the IPSL model is the closest to SOCAT observations ($\rho$ = 0.84 $\pm$ 0.02), although all models are reasonably close to each other, and therefore to SOCAT, with differences at the 1-2 $\sigma$ level, as shown in the bottom panel of Fig. \ref{fig:SOCATweights}.

\subsection{Transferability of representation from models to observations}
\label{sec:transfer_model_data}

\begin{figure}
    \centering
     \includegraphics[width = \linewidth]{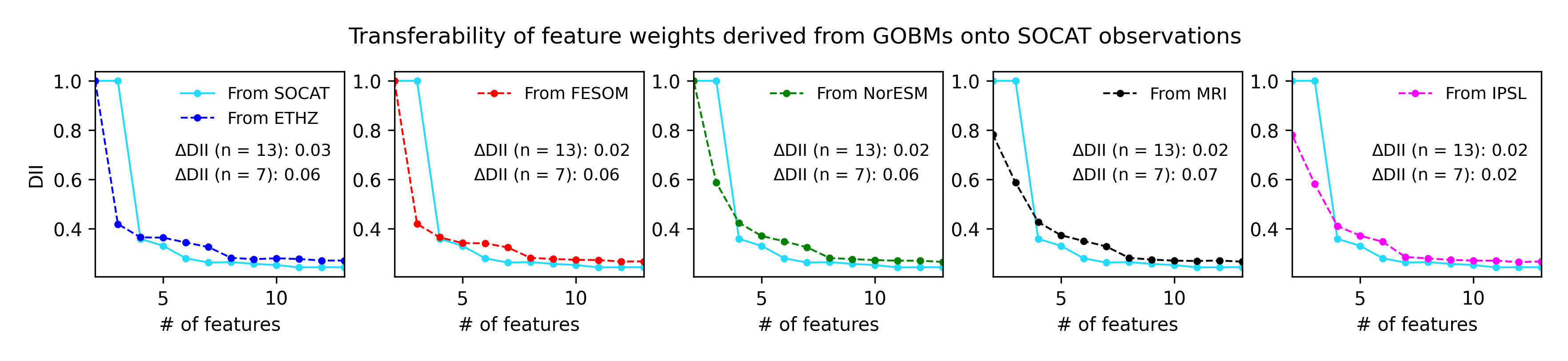}
    \caption{DIIs obtained by applying to SOCAT observations each GOBM's learned weights (dashed lines) and those earned on observations directly (solid line, same as in Fig. \ref{fig:ID_Socat_models} and in all panels). The weights learned in GOBMs spaces are highly predictive also in the full-size ($n$=13) observational space, as shown by the DII curves asymptotically approaching each other. However, using reduced-space representations is less optimal, as shown by the gaps for $n$ = 6-8.}
    \label{fig:modelweights-to-SOCAT}
\end{figure}

As we noted earlier in this section, if features are correlated, different sets of feature may be similarly effective in representing $\Delta$CO$_2$, because multiple combinations carry almost the same information. As a result, a model’s optimal feature set may differ from SOCAT’s but still be nearly as informative about $\Delta$CO$_2$. Figure \ref{fig:modelweights-to-SOCAT} tests this ``transferability of representations": we apply the optimal weights chosen by each model to SOCAT observations, compute the DII curve, and compare it to the one derived using the SOCAT observations directly. A higher DII would signify that we are learning a sub-obtimal representation by using models in place of observations. We note three main results (shown here for random seed 13 for clarity, but analog to what we find for the other random seeds):
\begin{enumerate}
    \item In the full space of input variables, the representation (i.e., the optimal feature set and their weights) learned from models is very predictive also in the observational space, as shown by the DII curves asymptotically approaching a value just slightly higher than the optimal curve that uses the weights derived from SOCAT observations (residual $\Delta$DII $\sim$ 0.02 for all models);
    \item The difference, however, is more marked for smaller feature sets ($n$ = 6 or $n$ = 7). Using the GOBMs only as a learning basis, we would have concluded that the relevant feature set has $n = 7$ variables (not incorrectly; $n = 6$ for observations), but the representation learned on that subset of 7 variables would yield a larger DII than the optimal one for most models ($\Delta$DII $\sim$ 0.06).  
    \item{Inter-model differences are small, although we note that for this specific random seed, IPSL shows the smallest gap for $n = 7$ ($\Delta$DII $\sim$ 0.02), indicating the closest match to SOCAT.}
\end{enumerate}

Because of the overall small differences between DII curves derived on GOBMs and observations, we conclude that optimizing feature representations in models is a generally valid approach for identifying informative features for real-world $\Delta$CO$_2$. This analysis also yields a second metric to compare representations: the vertical distance, at a given dimensionality of interest $n$, between the DII curve obtained by optimizing weights in the SOCAT observations space, and the DII curve obtained by using weights optimized in each GOBM and applied to the SOCAT space. For this use case, this metric would be the vertical distance between the light blue curve and each of the colored curves, for a chosen $n$, in Fig. \ref{fig:modelweights-to-SOCAT}.

\section{Effect of observations sparsity on generalization}
\label{sec:sparsity}

The distribution of observations in the SOCAT domain is significantly inhomogeneous, as can be seen e.g. in the left column of Fig. \ref{fig:resampling}. 
We can expect some biases in the representation learned from the SOCAT domain and applied to the full uniform space or other geographical domains; the scope of this section is to investigate their magnitude and behavior, first in models and then in observations.

\subsection{Full space vs SOCAT domain for GOBMs}

Gapless data on the whole global ocean only exists in GOBMs, not in observations; hence, we begin to study the generalization properties of the $\Delta$CO$_2$ representation using these models.
For all GOBMs, we compare two equally-sized ($n$ = 16,000) sets of points: the ``full space", which includes randomly selected points distributed homogeneously in space and time, and the ``SOCAT domain", where the distribution of points matches that of the SOCAT observations. Like before, we restrict the analysis to the 2020-2022 time frame, and we use  the intrinsic dimension (ID) and the differential information imbalance (DII) as tools to analyze different data spaces.

\begin{figure}
   \centering
       \includegraphics[width = \linewidth]{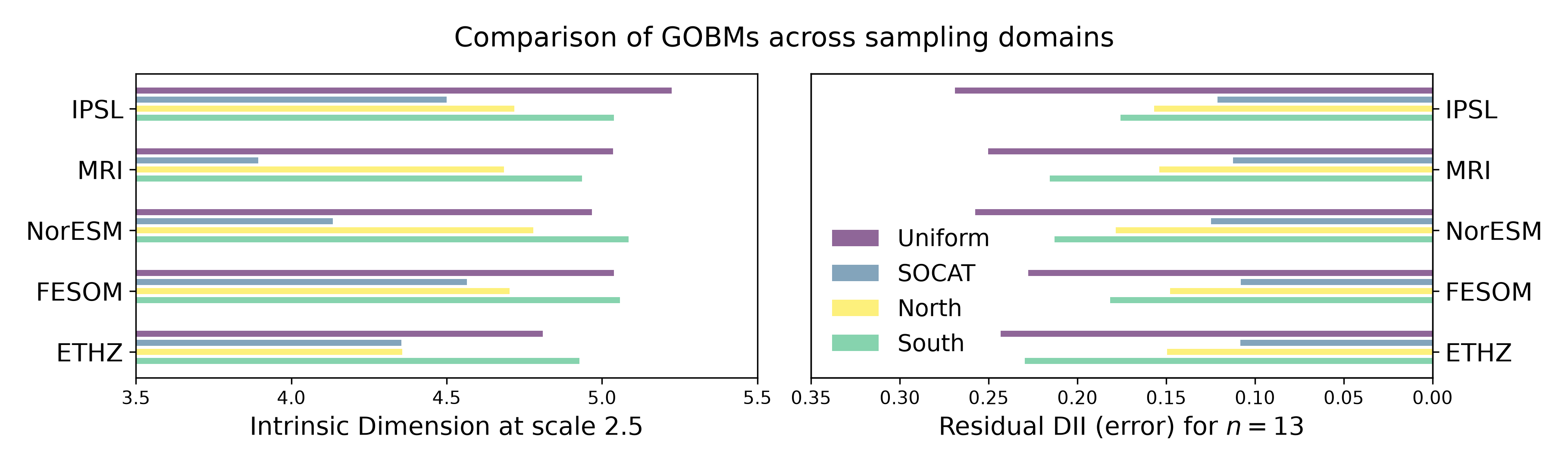}
    \caption{We show the intrinsic dimension, interpolated to scale 2.5 (see Figs. \ref{fig:ID_Socat_models} and \ref{fig:sup:GOBMsconv} for full curves for the SOCAT and full domain), in the left panel, and the DII in the right panel. In all GOBMs, the SOCAT space is significantly simpler (smaller ID) and the $\Delta$CO$_2$ field is easier to model (smaller DII) with the same input variables than the full model. Select latitude bands exhibit more variability across models, but for all GOBMs, the Southern Ocean band is more complex and less resolved than its northern counterpart. }
\label{fig:Domains}
\end{figure}

For all the models, we find that the full space is significantly more complex than the SOCAT domain. First, the intrinsic dimension (ID) is higher by 0.5-1 units (difference in purple vs blue bar in the left panel of Fig. \ref{fig:Domains}). 
Second, the residual DII is significantly higher in the full space (difference in purple vs blue bar in the right panel of Fig. \ref{fig:Domains}), indicating that the input variables can't predict the target in the full space to the same accuracy (DII is $0.12-0.14$ for the SOCAT domain, and $0.24-0.27$ in the full space). 

Nonetheless, we find that the representation of $\Delta$CO$_2$ learned within the SOCAT domain is very similar to the one learned from the full space. The weights assigned to each feature for different input space dimensionalities are highly correlated for all models, with consistently highest values for the NorESM model ($\rho \sim 0.95$ for $n$ = 13 and $\sim 0.95$ for $n$ = 7), the FESOM model ($\rho \sim 0.93$ for $n$ = 13 and $\sim 0.96$ for $n$ = 7), and the MRI model ($\rho \sim 0.92$ for $n$ = 13 and $\sim 0.9$ for $n$ = 7). We show the comparison of weights derived from the full domain and the SOCAT domain in Fig. \ref{supp:fig:WeightsSOCAT}, and the feature rankings for the full GOBM spaces in Table \ref{supp:tab:rankings}.

One difference in the feature importance that is consistent across models is that the (sine of) latitude $A$ is less important in the representation learned within the SOCAT domain, while the longitude becomes more important. Intuitively, SOCAT observations are less spread across different latitudes than those of the uniform domain and show significant value gaps in the southern oceans, which makes the variable $A$ less important within the SOCAT space.

\subsection{Stratification by latitude for GOBMs}

To further investigate the difference between the global ocean and specific spatio-temporal subsets in GOBMs, such as the SOCAT domain, we also assessed the predictive performance of our input features for $\Delta$CO$_2$ separately within northern and southern latitude bands. We restricted the spatial domain to $+30^\circ$ to $+60^\circ$ in the North and $-60^\circ$ to $-30^\circ$ in the South. Notably, the southern band contains significantly less land, resulting in more than twice as many data points than in the northern band.

As usual, we look at both ID and DII as diagnostic tools. 
Our study shows that the northern band tends to have lower intrinsic dimension than the full space, and a lower residual DII, indicating that $\Delta$CO$_2$ can be more accurately predicted from our 13 input variables in this Northern region. Results from this region are shown as yellow bars in Fig. \ref{fig:Domains}. In contrast, the intrinsic dimension of the Southern band is consistently higher, across all models, than that of the Northern band, and comparable with the global baseline. This suggests that the southern system is qualitatively different: given the same set of input variables, $\Delta$CO$_2$ is less well described in the South than in the North. The southern latitude band exhibits greater intrinsic complexity, both in the observations and in the GOBMs, likely attributable to the Southern Ocean's unique circulation and biogeochemistry \cite{Mongwe2018, Gruber2019, Crisp2023,Sauve2023}.

\subsection{Transferability from SOCAT domain to full space for GOBMs}

To further quantify the effect of the limited domain covered by SOCAT observations, we can also ask how accurate the representation learned from the SOCAT domain would be, if applied to the full ocean.

\begin{figure}
    \centering
    \includegraphics[width = \linewidth]{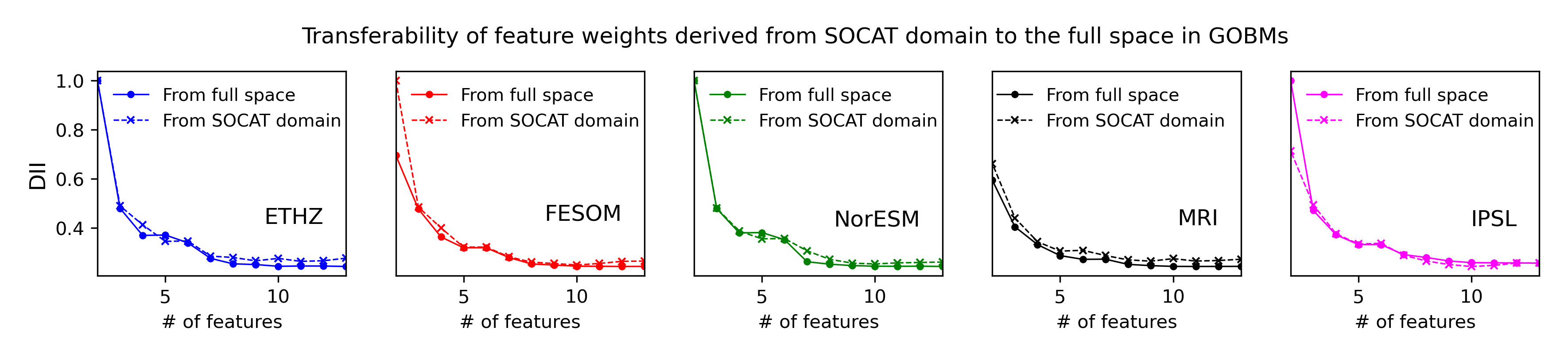}
    \caption{Dashed lines show the DIIs computed across the full geographical space of each GOBM, but using the features and weights that were optimized only within the SOCAT domain. Solid lines show the DIIs obtained derived from features and weights optimized in the full geographical space. Information loss during the transfer is shown by a higher DII for the dashed lines. We find that the two representations perform comparably well in the global ocean.}
    \label{fig:Generalization}
\end{figure}

To answer this question, we use the optimal weights derived from the SOCAT domain for the five GOBMs, we calculate the DII corresponding to this representation in the full domain, and we compare it with the DII calculated with the weights derived in the full domain. We find that the two DIIs are essentially the same (see Fig. \ref{fig:Generalization}), which means that the two representations are equally skilled at modeling $\Delta$CO$_2$. This is a non-trivial and encouraging result: it shows that \textit{at least in the GOBMs}, we can trust the representation learned from the SOCAT domain to be valid outside this domain and for more general tasks. We turn to the observations domain next. 

\subsection{Generalization tests for SOCAT observations} \label{sec:generalization_obs}

\begin{figure}
    \centering
    \includegraphics[width=\linewidth]{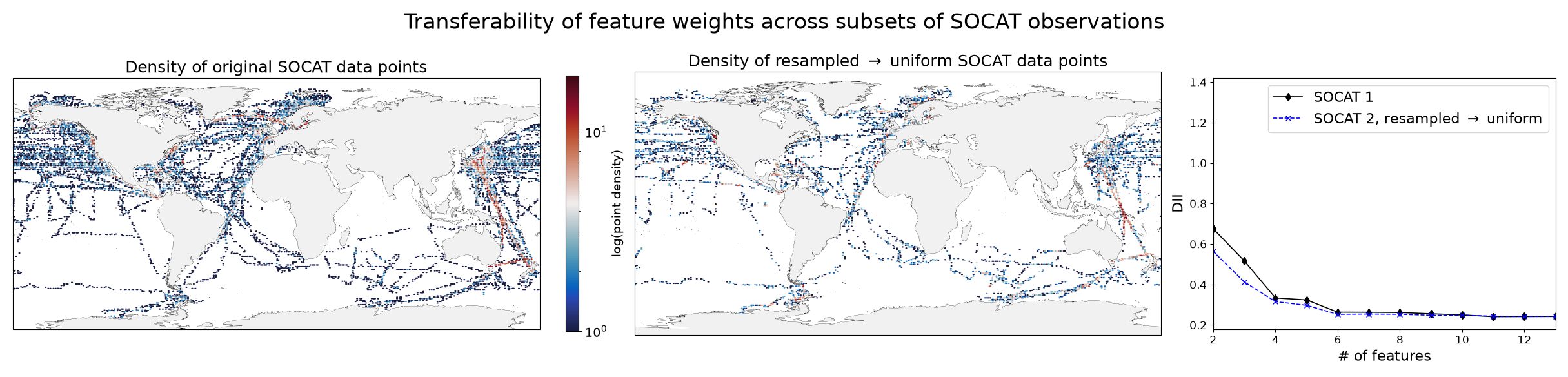}
    \includegraphics[width=\linewidth]{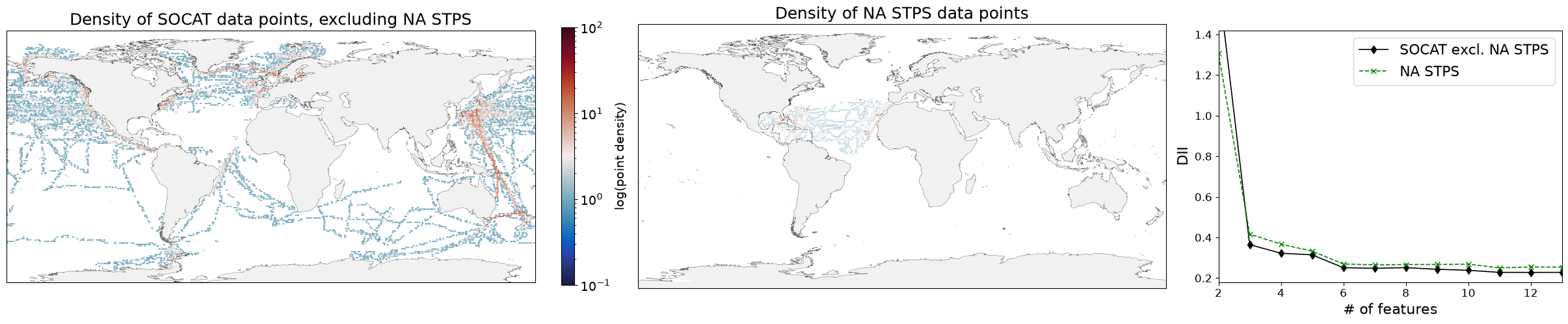}
    \includegraphics[width=\linewidth]{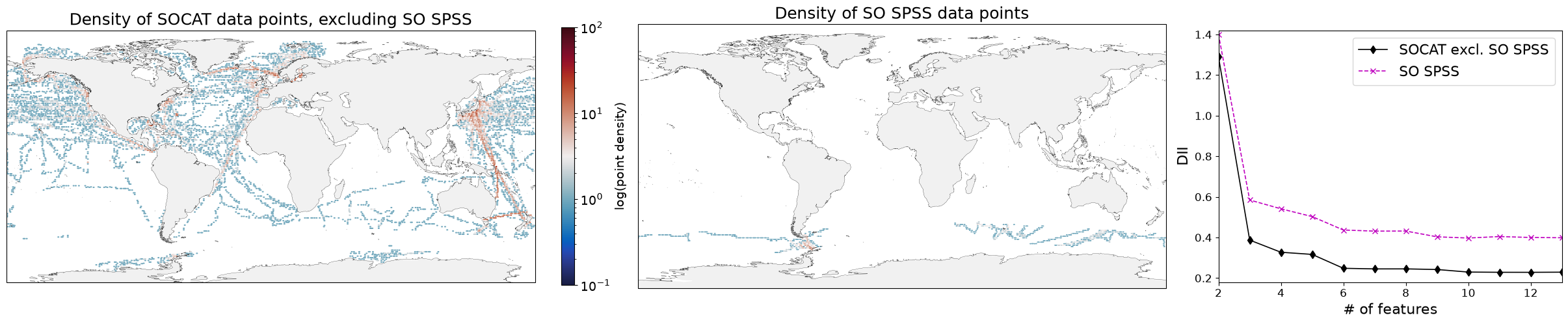}
    \caption{The three rows show how the representation of $\Delta$CO$_2$ learned in the first domain (left column) generalizes to the second domain (middle column), for observations from the SOCAT sample. The generalization is successful then the DIIs (right column) match - as in the first and second row - and unsuccessful then they don't (third row); the gap signals a mismatch in modeling across domains.}
    \label{fig:resampling}
\end{figure}

In this section, we aim to quantify the biases in learning the representation of $\Delta$CO$_2$ in the real ocean due to the sparsity and spatio-temporal bias of the SOCAT observations. Ideally, we would perform the same test we carried out in the previous section and compare the representation learned in the SOCAT domain to a representation learned via uniformly sampled observations, but the latter are not available. Therefore, we carry out three separate generalization tests. For each one, we derive the representation of $\Delta$CO$_2$ from data in an original domain, and we use it to derive the DII within that sample and for a different disjoint domain. If the two DIIs are similar, this indicates that the representation learned from domain 1 is equally skilled for domain 2 and the generalization is successful. If the DII for for domain 2 is significantly higher, the representation doesn't generalize successfully from domain 1 to domain 2. We show the domains for each test in the first two columns of Fig. \ref{fig:resampling}, and the DIIs in the third column.

\begin{enumerate}[leftmargin=7pt]
    \item In our first case study, we split the SOCAT observations for the reference time period (approximately 29,000 data points) in two distinct randomly sampled sets (SOCAT1 and SOCAT 2), and we use a chain of importance weighting steps to obtain a more uniformly sampled SOCAT2 by weighted resampling. In practice, we define a source (the subset of $A, B$ and $C$ feature distribution for the original SOCAT2, distributed like SOCAT) and a target distribution ($A, B$ and $C$ for a uniform distribution in space and time), and we carry out two steps. First, we use Unbalanced Optimal Transport \cite{chizat2018scaling} to derive a weight for each point, and then, we use IPF (Iterative Proportional Fitting) refinement \cite{deming1940least} to iteratively correct each of the (A, B, C) 1D-marginals and ``nudge" them towards their target distribution. The distance between source and target distribution is evaluated using the Sliced Wasserstein Distance \cite{bonneel2015sliced}, which is reduced from 0.395 to 0.109 in this process; more details about the resampling weights derivation are provided in the Appendices. The final product of this process is a set of resampling weights, and we obtain a new, ``more uniform SOCAT2" by sampling from the original SOCAT2 distribution using these weights as probabilities (so, data points in dense regions will be assigned a low resampling weight and many of them will not be selected in the new sample, while data points in sparsely populated regions may be sampled several times). 
    
    Our results are shown in the first row of Fig. \ref{fig:resampling} and indicate that the representation learned from SOCAT1 successfully generalizes to ``resampled towards uniform SOCAT2" (the weights derived on SOCAT1 yield essentially the same DII curve on uniform SOCAT2). This is encouraging, but we note that while this is a ``best case" scenario for obtaining a uniform sample from existing data that are distinct from SOCAT1 and fairly mimic the full ocean outside SOCAT1, the initial distribution is so skewed that the final sample is still not uniform. In particular, areas that are devoid of observations remain devoid of observations, as this is a resampling and not a generative procedure. We conclude that adding observations in undersampled areas, such as the Southern Ocean, is needed to strengthen this conclusion.

    \item In our second case study, we remove the North Atlantic Subtropical Permanently Stratified biome (NA STPS from \citeA{fay2014global}) from the SOCAT sample. We learn the optimal representation of $\Delta$CO$_2$ in this ``SOCAT excluding NA STPS" domain, and we compare the DIIs obtained with those weights for this domain and for the NA STPS biome. We find that the DIIs are very similar, indicating that the representation generalizes successfully.

    \item In our third case study, similarly to what we did in the previous one, we remove the Southern Ocean Subpolar Seasonally Stratified biome (SO SPSS from \citeA{fay2014global})from the SOCAT sample, and again we learn the representation of $\Delta$CO$_2$ in this ``SOCAT excluding SO SPSS" domain. We compare the DIIs for this domain and for the SO SPSS biome. Contrary to our finding in the previous case, we find that the DII in the subpolar Southern Ocean biome is significantly higher ($\sim$ 0.4 vs $\sim$ 0.23 in the full 13-dimensional space), indicating that the representation learned in the broader domain that excludes data from SO SPSS doesn't correctly capture $\Delta$CO$_2$ in this region. Once again, our findings point to the need for additional measurements, particularly in the sparsely observed Southern Ocean \cite{Gloege2021,Hauck2023,Heimdal2024,HeimdalMcKinley2024}.
    \end{enumerate}

\subsection{Robustness across time periods}
\label{sec:time_biomes}

\begin{figure}[h!]
    \centering
    \includegraphics[width=\linewidth]{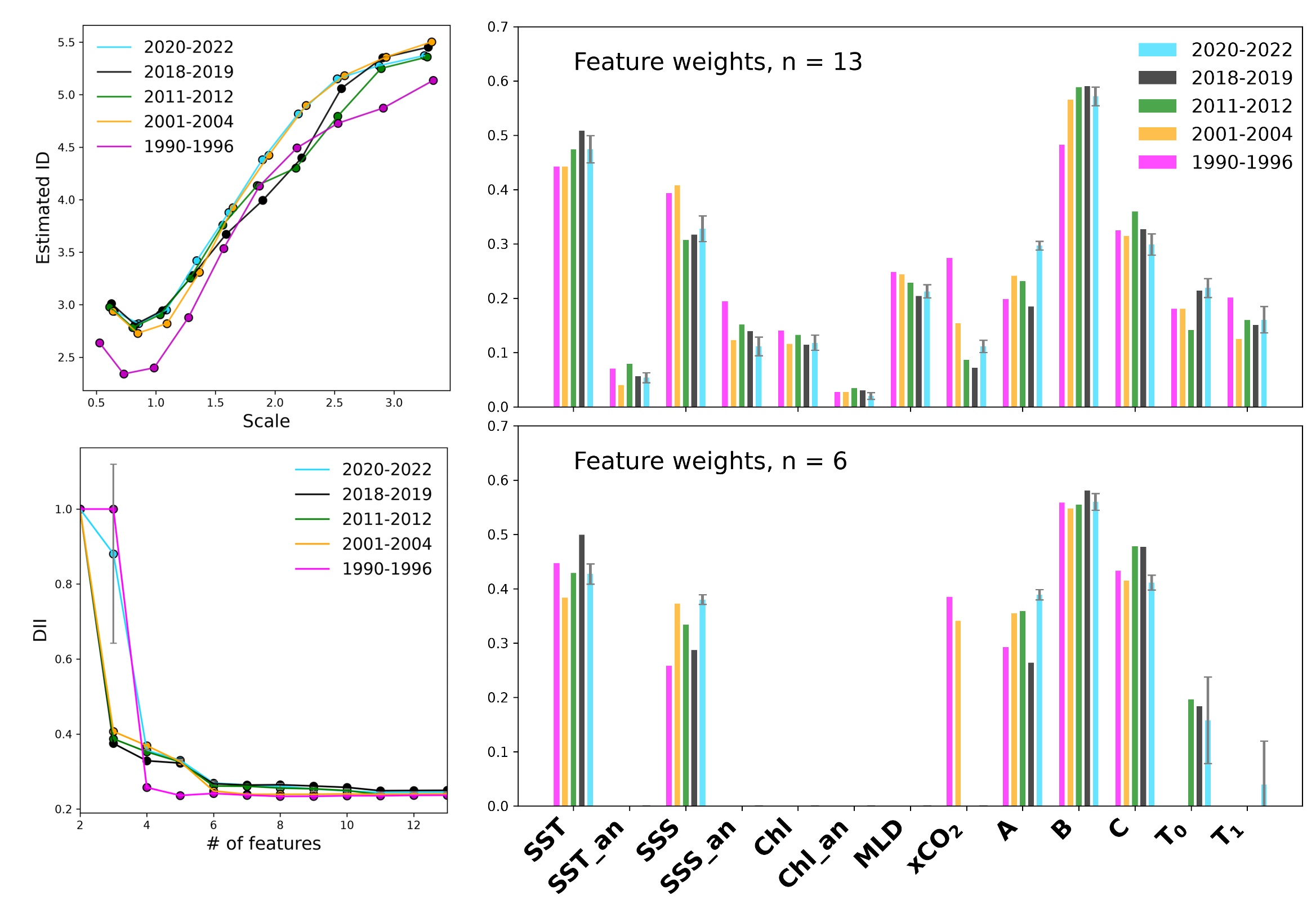}
    \caption{Comparison of intrinsic dimension (top left), DII (bottom left), and feature weights (right) for n = 16,000 randomly selected SOCAT observations in five time periods. Error bars coming from sample variance are included for the reference period (2020-2022).}
    \label{fig:acroSSTime}
\end{figure}

Our analysis so far has focused on SOCAT observations (or equivalently selected GOBMs data points) in the three-year period between January 2020 and December 2022. To understand how the choice of time period affects the learned representation, we repeated the analysis of Sec. \ref{sec:SOCAT_only} for four equally sized ($n$ = 16,000) random selection of SOCAT observations across time, starting in the 1990s. The time intervals were selected to be sufficiently spaced apart to detect evolution, and to contain a similar number of data points to keep the sampling rate approximately constant. Our results are shown in Fig. \ref{fig:acroSSTime}. 

We observe that the complexity of data space, estimated through the ID of the five data manifolds, and the predictive capability of input features, assessed through the DII curve, remain approximately constant across time after the year 2000. A plateau at $n = 6$, indicating that six features capture the vast majority of information about the target $\Delta$CO$_2$, is present in the same time frames. The 1990-1996 data manifold exhibits some differences from others, with an estimated ID lower by $\sim$ 0.5, and a plateau in the DII curve at $n = 4$. We note that the SOCAT database contains significantly fewer data points in the 1990s, with an average of $\sim$ 4300/year for the 1990-1996 time period, a factor of 4-5 smaller than in the 2010s. Such extreme sparsity can bias estimates from SOCAT-based products \cite{Dong2024} and explain the lower complexity.

Feature importances are also in good agreement across time periods, with the exception of atmospheric CO2 ($xCO_2$), which is more important in the 1990-1996 and 2001-2004 time frames. This may originate from data sparsity or from the definition of the target variable as $\Delta$CO2, which is expected to lower the importance of the $xCO_2$ feature as long as the time dependence of ocean CO$_2$ and atmospheric CO$_2$ is the same. The higher importance of $xCO_2$ may indicate that such hypothesis is not correct in those time frames.

\section{Using the optimal distance metric in machine learning models}
\label{sec:ML}
In this section, we want to show how the optimal representations (i.e., the features and weights in each dimension) found through the information imbalance formalism can be used to improve the outcomes of distance-based machine learning models that use the same set of features and target. We use the k Nearest Neighbor algorithm \cite{cover1967nearest} for this test, a simple algorithm that predicts the $\Delta$CO$_2$ of any point by looking at the $k$ nearest neighbors of that point in the feature space that are present in the training set, and reporting a simple or weighted average of their $\Delta$CO$_2$. In our case, an optimization procedure led to selecting $k = 5$ as the number of neighbors, and a weighted mean with weights inversely proportional to the distance of each neighbor to the point under consideration. These hyperparameters were robust to different GOBMs and data configurations.

We look at 1) the error made by predictive models trained on a random subset of SOCAT observations (the training set) when applied to a disjoint but statistically equivalent subset (the test set), and 2) the error made by predictive models applied to the same random subset of SOCAT observations as before, but with feature weights learned using the GOBMs. All scores and errors are calculated as the average of 5-fold cross-validation process, and denoted as ``test error" or ``test score" to clarify that these sets have not participated in the training or parameter tuning process.

\begin{figure}
    \centering
    \includegraphics[width=0.38\linewidth]{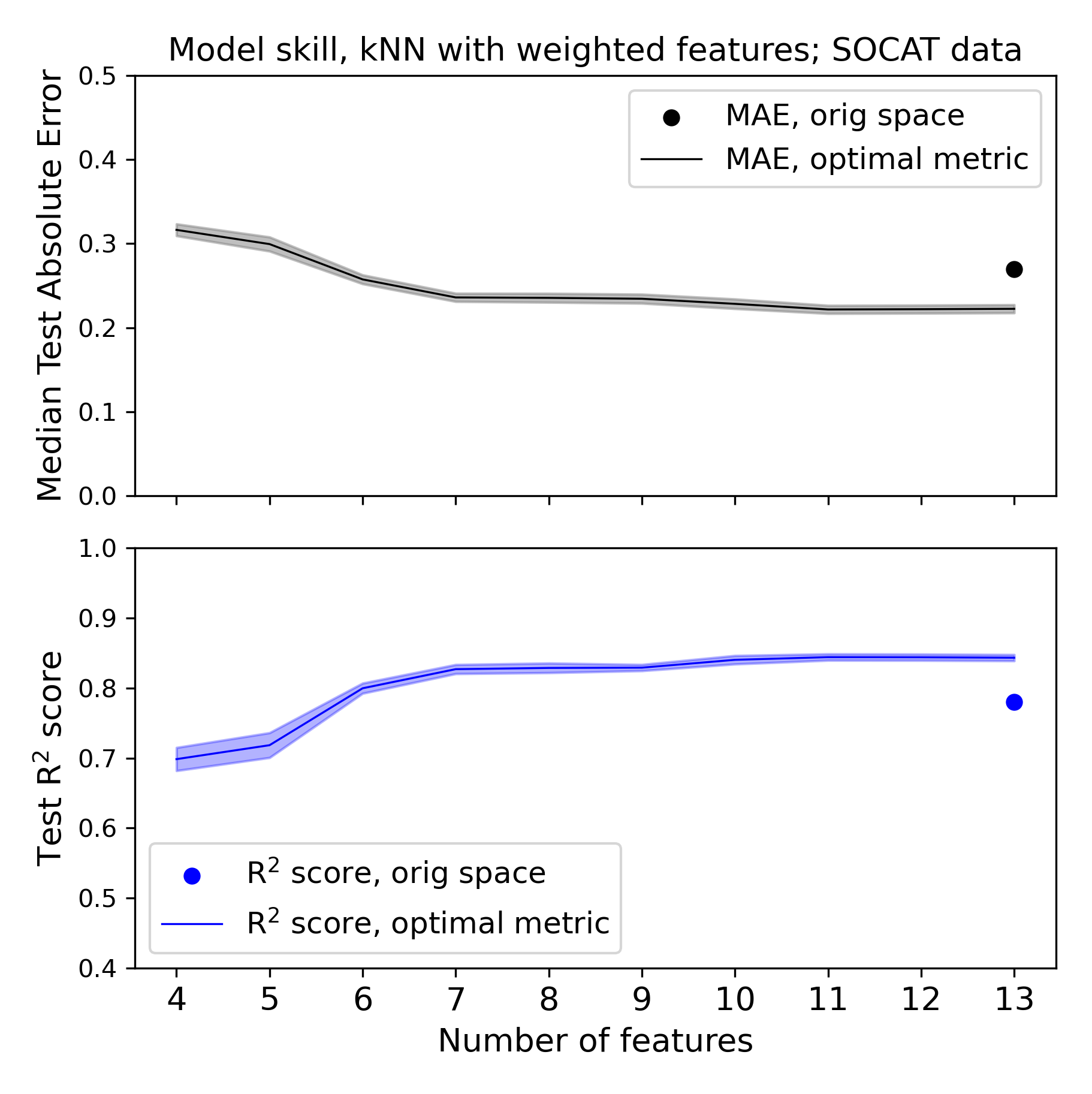}
        \includegraphics[width=0.5\linewidth]{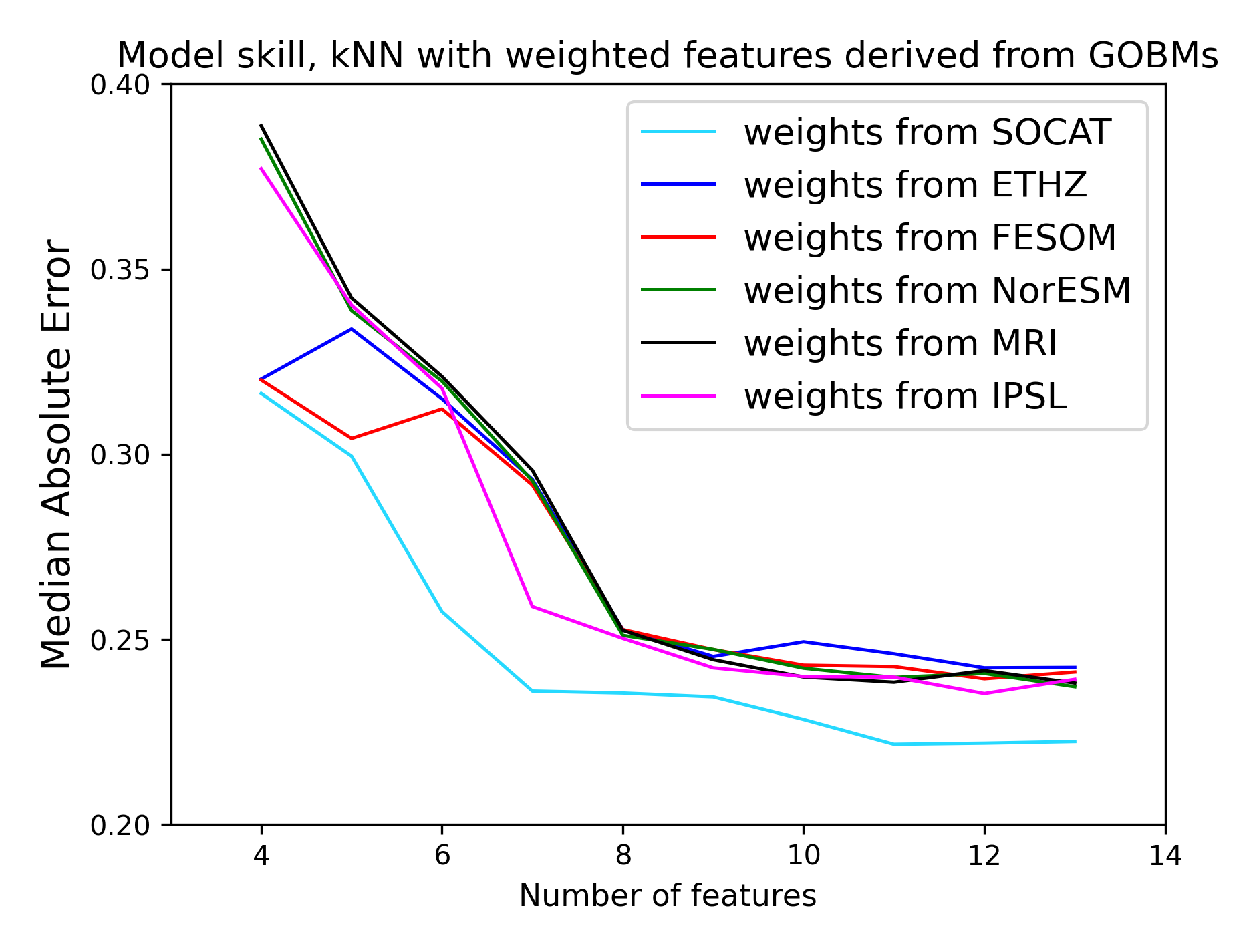}
    \caption{\textit{Left panel}: Median absolute error (top) and R$^2$ score (bottom) for the kNN machine learning model applied to a test set of SOCAT observations, with unweighted features (dots), and with feature weights derived through DII optimization, for each $n$ (lines). The error (score) decreases (increases) noticeably by using the optimized representation, and it holds stable for smaller spaces as well. \textit{Right panel}: Test error for optimized feature sets derived directly from SOCAT observations (light blue line), or derived from GOBMs and applied to SOCAT observations (other colors). The behavior of the error closely follows the DII curves obtained in the ``Transferability of weights" section.}
    \label{fig:ML}
\end{figure}

For our first test, we find that using the optimal representation (i.e., scaling the features by the weights found by the DII optimization in each dimension) before applying the $k$NN algorithm to the SOCAT observations in the usual time period 2020-2022 leads to a notable improvement in model skill for $n = 13$, shown in the left panel of Fig. \ref{fig:ML}. The median absolute test error, MAE, decreases from 0.27 to 0.22, while the $R^2$ score, a measure of correlation between predicted and true values, increases from 0.78 to 0.84. Furthermore, both metrics remain stable even when smaller feature sets are considered, down to $n = 6-7$, coherently with the behavior of the DII, which remains essentially flat to the same extent, as shown in panel ``b" of Fig. \ref{fig:SOCAT_only}. 

In the right panel of the same figure, we show the median absolute test error of each $k$NN model applied to SOCAT observations, but with weights derived either directly from the SOCAT observations, or from each GOBM. This is an applied version of the ``weights transferability" test carried out in Sec. \ref{sec:transfer_model_data} and shown in Fig. \ref{fig:modelweights-to-SOCAT}, and we use the same random seed. The model skill decreases a bit (MAE increases from 0.22 to 0.24) when the GOBM-learned weights are applied instead of the native (SOCAT-learned) weights, but most importantly, the pattern of the error follows closely the DII curve found for this same transferability test (for example, IPSL is significantly closer than other GOBMs to the SOCAT curve for $n = 7$) in Sec. \ref{sec:transfer_model_data}. This result reinforces the idea that the vertical distance in the DII curve obtained with weights derived in the observations space and the DII curve obtained with weights derived using the GOBMs is a useful metric to compare models and observations.

\section{Conclusions}

In this work, we used two statistical tools, the intrinsic dimensionality (ID) and the Differentiable Information Imbalance (DII), to characterize and compare how surface ocean carbon ($\Delta$CO$_2$) can be represented by other features in SOCAT observations and in five widely used Global Ocean Biogeochemical Models (GOBMs). Our approach is model-independent, interpretable, and grounded in the geometry of the data manifold rather than in the performance of any particular predictive algorithm. Our main findings are below.

Our first finding is that the space of SOCAT observations for the 2020-2022 time period has an intrinsic dimension ($\text{ID} \sim 5.5$), which means that 5.5 perfect (but not necessarily physical) orthogonal features could capture the full information. We find that six features (salinity, sea surface temperature, and spatio-temporal features A, B, C, and T$_0$) carry the majority of information, and that the residual information imbalance (DII) between input features and target is $0.25-0.27$, i.e., closest neighbors in feature space are typically found within the first 12-14 percentiles of neighbors in $\Delta$CO$_2$ space. Together, these results suggest that {\bf the current set of features is a good, although not perfect, set of predictors for $\Delta$CO$_2$}. 

Next, we found that the representation of $\Delta$CO$_2$ in GOBMs is simplified with respect to observations: $\text{ID} \sim 4.5$ for any model, lower by about one unit compared to SOCAT space. This means the real-world data cloud varies along directions in feature space that models do not reproduce. Consistently, the DII between input variables and $\Delta$CO$_2$ is substantially lower in models (${\sim}0.10$–$0.12$) compared to observations.
The complexity gap between models and observations observed both in ID and DII suggests that GOBMs may underestimate the richness of the ocean carbon system, with potential implications for the accuracy of ocean carbon sink estimates in the Global Carbon Budget. Optimal feature weights derived from the models are similar to those derived directly from observations for some features, but all models overestimate the importance of chlorophyll and salinity anomaly, and underestimate that of spatial feature B with respect to SOCAT data (see Fig. \ref{fig:SOCATweights}). The IPSL model is the closest to observations for this time period. Despite these differences, we find that feature weights and sets derived from models perform nearly equally well in predicting $\Delta$CO$_2$ in observations as feature weights and sets derived from the observations themselves. Therefore, \textbf{all five GOBMs are effective proxies for the real ocean, at least in the SOCAT domain, even if they are less complex}. 

Third, we examined the role of observational sparsity in the learned representations. In GOBMs, where we can consider any spatio-temporal domain, we find that the representation of $\Delta$CO$_2$ in the SOCAT domain is significantly simpler (lower ID, lower residual DII) than in a global, uniformly distributed domain. Nonetheless, the feature weights remain consistent between the full space and and SOCAT space (typical correlations between weight vectors are $\sim$ 0.9, see Fig. \ref{supp:fig:WeightsSOCAT}), and the representation learned within SOCAT performs comparably well to the optimally selected one, when applied to the full domain (Fig. \ref{fig:Generalization}). This is an encouraging result for the machine learning community building data-driven CO$_2$ products, but we also note that each ML algorithm will have a different learning profile, and especially ML models that don't use the notion of metric space, such as tree-based method, may behave differently in generalization.

Motivated by this finding, we also explored how the ID and DII change across regions in GOBMs, and found that spatially restricting the analysis to northern or southern latitude bands improves predictivity (Fig. \ref{fig:Domains}), especially in the northern band, and inter-model variability is generally lower than inter-region variability. These results confirm that the $\Delta$CO$_2$ may vary across regions in a way that the current representation cannot capture, which may fundamentally limit the performance in a global analysis. 

We also tested the transferability of learned representations within SOCAT observations, within the limits of existing data, which are biased and sparse.
Those revealed an important asymmetry. The representation learned excluding the North Atlantic transfers successfully to that region, but the same does not hold for the Southern Ocean: withholding it from training causes the DII to rise markedly from ${\sim}0.23$ to ${\sim}0.40$, signaling the poor quality of the learned representation in this region. This quantitatively confirms that the Southern Ocean is not merely undersampled, but dynamically distinct from other regions. The combination of fewer data and a more difficult target makes the southern latitudes a clear priority for expanded data collection and, potentially, the development of dedicated models. 

We examined the relationship between input features and target across different periods of time, selecting similarly-sized data set across the 1990s, 2000s, and 2010s in addition of our reference 2020-2022 period, finding largely consistent manifold dimensionality, feature weights, and DII curve from the 2000s onwards, while the early data sets, in particular the 1990-1996, exhibited distinct characteristics with lower ID and a smaller set of informative features.

Our results also yield implications for machine learning based gap-filled reconstructions of ocean carbon. We demonstrated that the optimal feature weights derived from the DII can be used directly to improve distance-based machine learning algorithms. Replacing the Euclidean distance with our DII-optimized metric yields a ${\sim}20\%$ improvement across all evaluation metrics in a $k$-Nearest Neighbor model of $\Delta$CO$_2$. This result has direct implications for many ML-based CO$_2$ reconstruction methods underpinning current data products, and suggests that manifold-informed feature weighting could become a standard component of such pipelines. 

Finally, beyond the individual findings above, this work also established a new general, quantitative framework for evaluating GOBMs or Earth System Models against observations or against each other without relying on any specific predictive model. In particular, we proposed two metrics to compare the representation of a target variable in different data spaces: 1) the correlation coefficient between optimal weights vectors derived by the DII in each of the two spaces, considered for the smallest maximally informative dimension of such spaces; and 2) the vertical distance, at a given dimensionality of interest $n$, between the DII curve obtained by optimizing weights in the first space, and the DII curve obtained by using weights optimized for the second space and applying such representations to data from the first space. When applied to the comparison of $\Delta$CO$_2$ representation in GOBMs and observations, we found that all GOBMs are similarly close to observations, with IPSL faring better than others at the $1.5\sigma$ level in both metrics. We hope that this evaluation framework will be valuable for GOBM developers and the Global Carbon Budget community alike, both as a benchmarking tool and to define a evidence-based weights when averaging different models. 

Taken together, our results reveal a fundamental and previously unquantified complexity gap between the real ocean carbon system and its model representations, identify the Southern Ocean's dynamically distinct character as a key priority for both model improvement and targeted observational campaigns, and provide the community with a portable, interpretable toolkit for the ongoing evaluation of GOBMs and machine learning-based carbon products.

\appendix

\section{Notation and variable setup}

We summarize the terminology used throughout the paper in Table \ref{supp:tab:notation}, and summarize the input features and the target variable in Table \ref{supp:tab:variables}.

\renewcommand{\arraystretch}{1.2}

\begin{longtable}{p{0.28\textwidth}p{0.67\textwidth}}
\caption{Glossary of terms used throughout the paper.}
\label{supp:tab:notation}\\
\hline
\textbf{Term} & \textbf{Definition} \\
\hline
\endfirsthead

\hline
\textbf{Term} & \textbf{Definition} \\
\hline
\endhead

Input or Feature space &
The multidimensional space defined by the input features used to estimate the target variable. \\

Target space &
The space defined by the target variable to be estimated. In this work, $\Delta$CO$_2$. \\

Scale &
Level of ``zoom,'' expressed in units of the typical distance between neighboring points. \\

Neighbors &
Data points that are close in a given metric space (e.g., feature space or target space). \\

Neighbor rank ($r_{ij}$) &
Ordinal position of point $j$ in the ordered list of neighbors of point $i$. The closest neighbor has rank 1, the second closest has rank 2, and so on. \\

Intrinsic dimension (ID) &
Minimum number of orthogonal variables required to describe the structure of a data manifold. They may not coincide with physical variables. \\

ID plateau &
Region where the estimated ID remains approximately constant as the scale increases. When present, it identifies the ID of the data manifold. \\

Optimal feature weights &
Feature weights obtained through an optimization process by maximizing the alignment between neighbor ranks in input space and in target space. \\

Feature ranking &
Ordering of the input features according to their optimal weights, from most to least informative for estimating the target variable. \\

Differential Information Imbalance (DII) &
Minimum error associated with estimating the target variable from the optimally weighted input features for a given dimensionality. \\

DII curve &
DII as a function of the number of retained features, from $n=1$ to the full feature space (here $n=13$). Often shown for $n \geq 3$, as lower-dimensional spaces can exhibit numerical instabilities. \\

Residual DII &
DII of the full feature space (here $n=13$), representing the irreducible error associated with the available feature set. \\

Smallest maximally informative space &
The smallest feature subspace for which adding additional features does not produce a meaningful reduction in DII. It is identified by a plateau or a pronounced elbow in the DII curve. \\

\hline

\end{longtable}

\begin{table}[h!]
\centering
\caption{Summary of feature and target variables; inspired by \cite{bennington2022}.}
\label{supp:tab:variables}

\begingroup
\setlength{\tabcolsep}{9pt}      
\renewcommand{\arraystretch}{1.2} 
\begin{tabular}{llll}
\hline
\textbf{Variable} & \textbf{Abbrev.} & \textbf{Transformation} & \textbf{Notes} \\
\hline
Sea surface temperature & SST & -- & \\
SST anomaly & SST$_{an}$ & SST $-$ monthly climatology & \\
Salinity & SSS & -- & \\
SSS anomaly & SSS$_{an}$ & SSS $-$ monthly climatology & \\
Atmospheric CO$_2$ & $x$CO$_2$ & -- & \\
Chlorophyll-$a$ & Chl & $\log_{10}(\mathrm{Chl})$ & \\
Chlorophyll-$a$ anomaly & Chl$_{an}$ & Chl $-$ monthly climatology & \\
Mixed layer depth & MLD & $\log_{10}(\mathrm{MLD})$ & \\[2mm]

\multirow{3}{*}{Geographic location}
& A & $\sin(\lambda)$ 
& \multirow{3}{*}{\makecell[l]{$\lambda$ = latitude\\$\mu$ = longitude}} \\

& B & $\sin(\mu)\cos(\lambda)$ & \\

& C & $-\cos(\mu)\cos(\lambda)$ & \\[2mm]

\multirow{2}{*}{Time of year}
& T$_0$ & $\cos(2\pi j/365)$ 
& \multirow{2}{*}{$j$ = day of year} \\

& T$_1$ & $\sin(2\pi j/365)$ & \\
\hline
\textbf{Target} & \textbf{Abbrev.} & \textbf{Transformation} & \\
\hline
$\Delta$CO$_2$ & -- & fCO$_2 - x$CO$_2$ & \\
\hline
\end{tabular}
\endgroup
\end{table}

\section{Convergence tests}\label{sec:Convergence}

We assessed the convergence of the ID, optimal weights, and DII for two important case studies representing observations and models and with very different sample sizes and sampling fractions. 

\begin{figure}[h!]
    \centering
    \includegraphics[width=\linewidth]{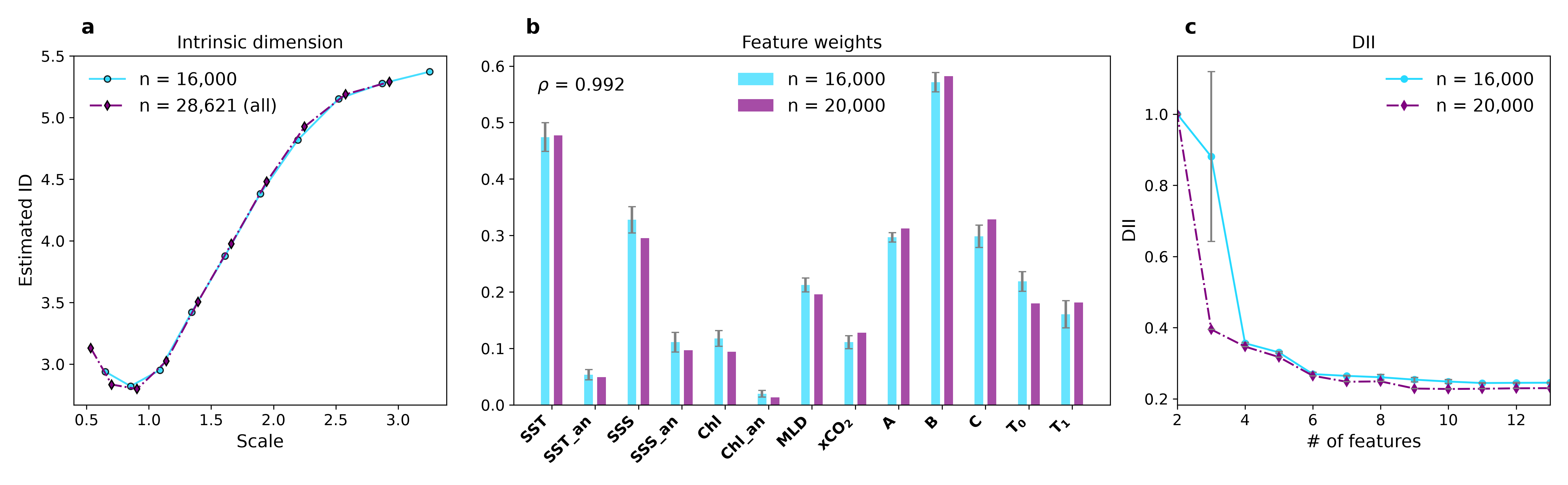}
    \caption{ID, optimal feature weights, and DII curve for SOCAT observations for different sample sizes. For the ID, where calculations can be easily run for larger sample size, we compare $n = 16,000$ (used in the paper) and the full sample ($n = 28,621$). For the DII and weights, we compare $n = 16,000$ and $n = 20,000$.}
    \label{fig:sup:SOCATconv}
\end{figure}

The first one is for SOCAT observations in the reference period (2020-2022), which contains 28,621 data points. The ID can be easily calculated even for large data set sizes, so we compared the ID for the full space with the one derived for our chosen sample size $n = 16,000$. We find that the IDs estimated with 16,000 and 28,621 are essentially indistinguishable from each other (panel a of Fig. \ref{fig:sup:SOCATconv}). 

The calculation of the DII requires a large amount of RAM and cannot currently be parallelized, so we worked with 20,000 points for comparison.  The weights derived for the two sample sizes are very close, with a Pearson correlation $\rho > 0.99$. The DII curves are also within 5\% of each other. It is possible that adding more points would lower the DII curve by a few percent points, but since we only use the DII for a comparison among equally-sized samples (for example, the DII from SOCAT observations and the DII from GOBMs within the same spatio-temporal domain), we conclude that our results are robust. 

\begin{figure}[h!]
    \centering
    \includegraphics[width=\linewidth]{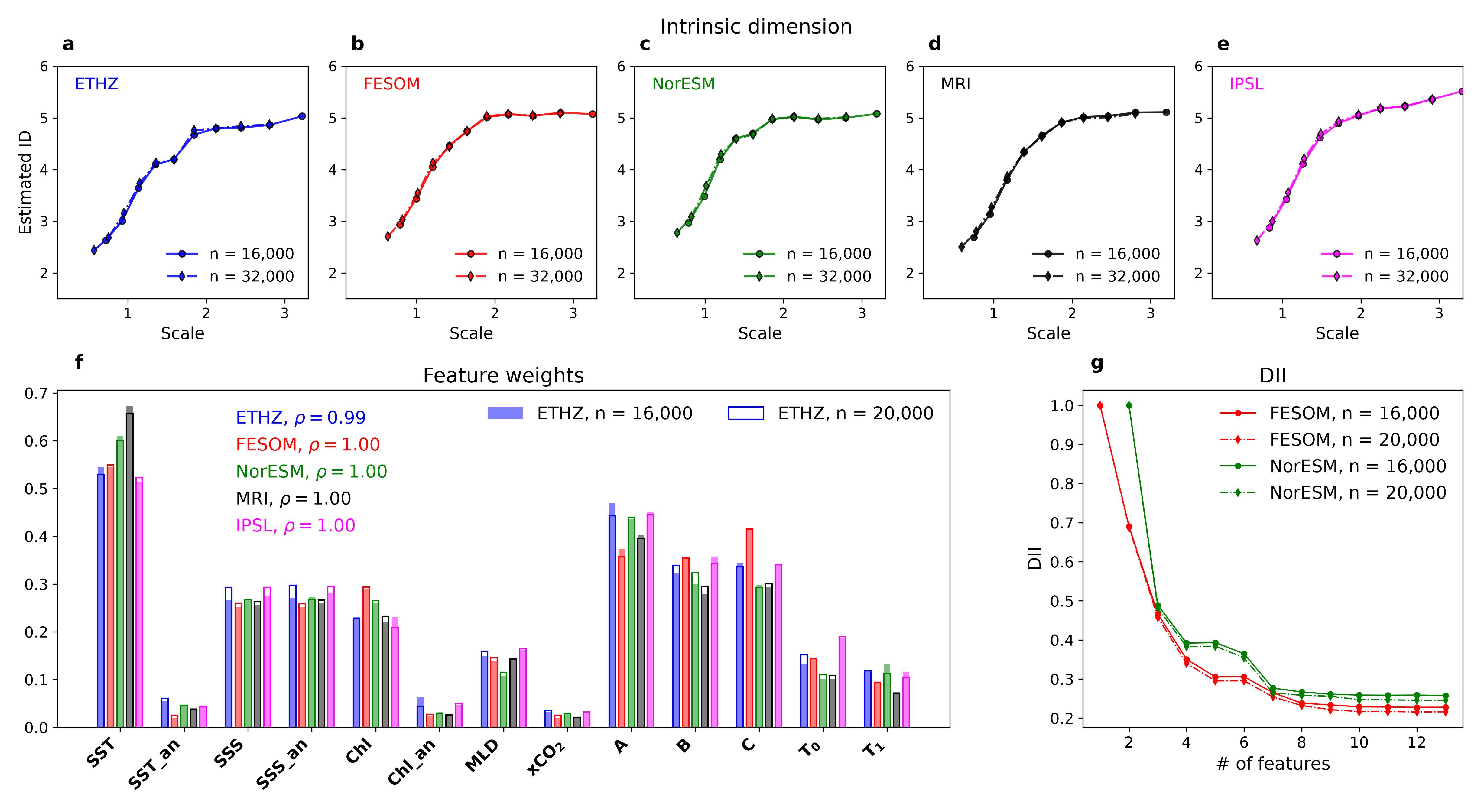}
    \caption{ID, optimal feature weights, and DII curve for the full uniform space in each of the five GOBMs for different sample sizes. For the ID, where calculations can be easily run for larger sample size, we compare $n = 16,000$ (used in the paper) and $n = 32,000$. For the DII and weights, we compare $n = 16,000$ and $n = 20,000$. We show DII curves for two models only to avoid overcrowding the plot, but the behavior is very similar for all five GOBMs.}
    \label{fig:sup:GOBMsconv}
\end{figure}

We also ran similar tests for the full, uniform domain in GOBMs that is used in Sec. \ref{sec:sparsity}. This data set is much larger than the previous one and contains 2,332,800 data points, so we are exploring much smaller sampling fraction by selecting $n = 16,000$ points. However, because points here are uniformly distributed through space and time, the sampling fraction required to learn their statistical properties will be much smaller. For each GOBMs, we estimate the ID using $n = 16,000$ and  $n = 32,000$, again finding that they are essentially indistinguishable from each other (panels a-e of Fig. \ref{fig:sup:GOBMsconv}). We note that the ID of such data spaces is similar to the one of the (much smaller) SOCAT observations, and therefore it makes sense that they would be correctly captured by a similar number of samples. We also compare the optimal weights and the DII curves found, for each model, for $n = 16,000$ and $n = 20,000$ samples. We find that weights are very close, with correlations $\rho \geq 0.99$ for all models (panel f of Fig. \ref{fig:sup:GOBMsconv}), and DII curves are also within a few percent of each other (panel g). Our conclusions is that these results are generally reliable, and that a small statistical error (of the order of 5\%) may only be observed when comparing the DII curves for very different sample sizes.

\section{Comparison of GOBMs and observations in the SOCAT domain}

We report here the ranking of features derived from the optimization of the DII, for SOCAT observations and for each GOBM in the same domain and using the same random seed. At each stage, starting with $n = 13$ and acting recursively, optimal weights for the $n$ features are calculated by maximizing the degree of similarity between neighbor ranks in input space and neighbor ranks in target space. Then, the feature with the lowest weight is eliminated and the process is repeated for the space with dimensions $n - 1$. The table shows that the least important features (xCO$_2$, Chl$_{an}$, and SST$_{an}$) and the most important features (SST, A, B, C) are common to the observations and to most models, while many differences emerge in the ``middle tier" rankings.

\begin{table}[h]
\footnotesize
\setlength{\tabcolsep}{4pt}
\begin{tabular}{l|ccccccccccc}
\hline
 Reference &  1, 2, 3 & 4 & 5 & 6 & 7 & 8 & 9 & 10 & 11 & 12 & 13\\
 \hline
 SOCAT obs & SST, B, C & A & SSS & T$_0$ & T$_1$ & MLD &  SSS$_{an}$& Chl & xCO$_2$ & SST$_{an}$& Chl$_{an}$ \\
 \hline
 ETHZ  & SST, A, C & B & SSS$_{an}$& SSS & Chl & T$0$ & MLD & SST$_{an}$& T$_1$ & xCO$_2$ & Chl$_{an}$\\ 
     \hline 
  FESOM & SST, A, C & B & SSS & SSS$_{an}$& Chl & T$_0$ & T$_1$ &MLD & SST$_{an}$& xCO$_2$ & Chl$_{an}$ \\
     \hline 
     NorESM & SST, SSS$_{an}$, A & C & B & SSS  & Chl & T$_0$ & T$_1$ & MLD & SST$_{an}$&  xCO$_2$ & Chl$_{an}$ \\
     \hline 
     MRI & SST, SSS$_{an}$, A & C & B & SSS & Chl & T$_0$ & MLD & T$_1$ & SST$_{an}$ & Chl$_{an}$& xCO$_2$ \\ 
     \hline 
     IPSL & SST, SSS$_{an}$, A & B & C & SSS & T$_0$ & Chl & MLD & T$_1$ & SST$_{an}$& xCO$_2$ & Chl$_{an}$ \\
     \hline
\end{tabular}
\caption{The table shows the ranking of features progressively eliminated during the optimal DII search as a function of $n$, for SOCAT observations and for each GOBM in the same domain. Column 13 shows the least important feature for $n$ = 13, column 12 shows the least important feature for $n$ = 12, and so on. The smallest size we consider is $n$ = 3. The sets of most and least important features are generally consistent, but there are important differences in the middle rankings.} 
\label{supp:tab:SOCATrankings}
\end{table}

\section{Comparison of feature weights derived from SOCAT domain and full space in GOBMs}

\begin{figure}
    \centering
    \includegraphics[width = 0.9\linewidth]{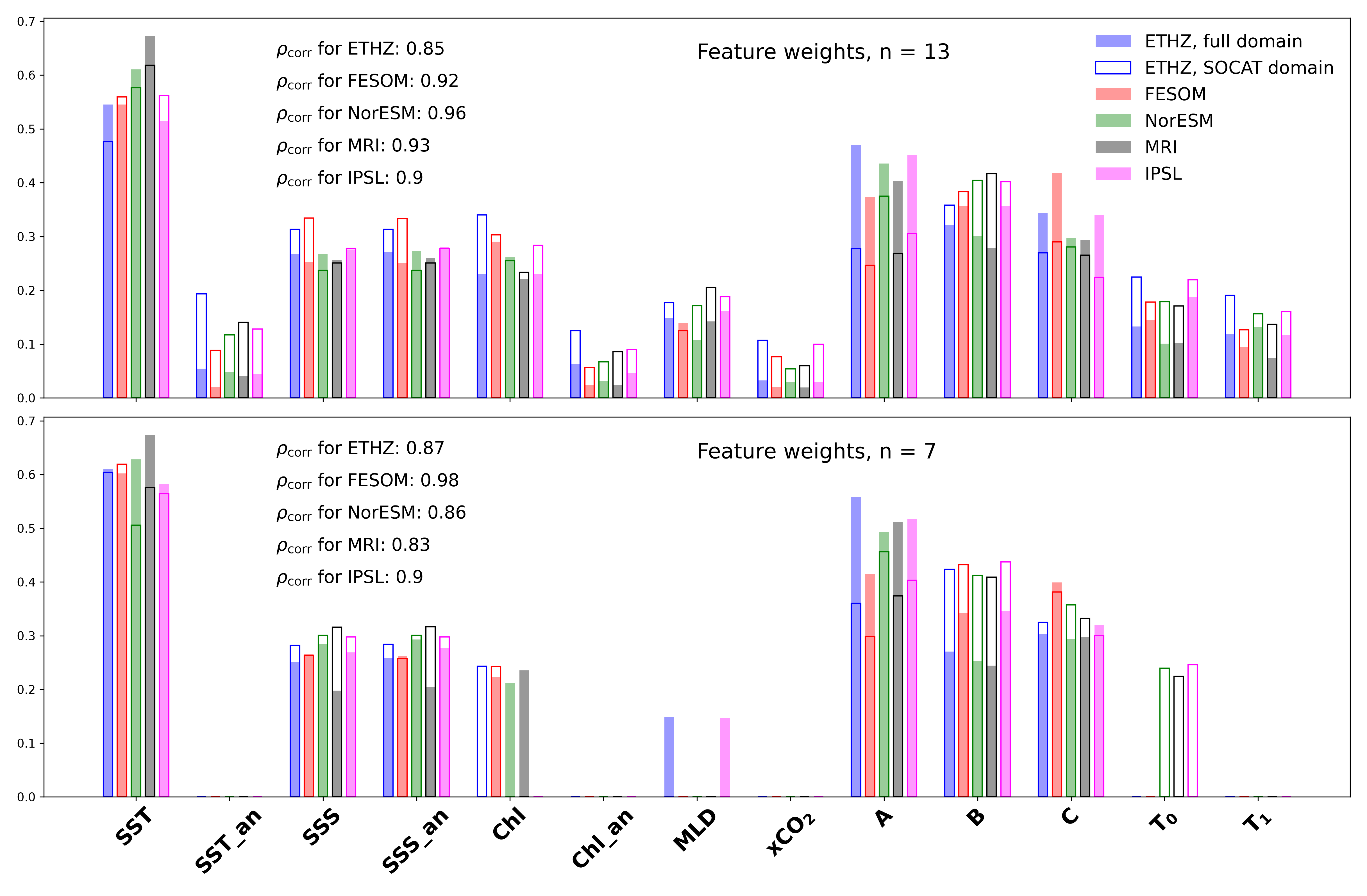}
    \caption{Feature weights derived by optimizing the DII for the GOBM models, using the full space where points are uniformly distributed in space and time (solid bars) and the SOCAT observations mask (hollow bars), for $n = 13$ (full space) and $n = 7$ (smallest fully informative space). The representations obtained from the two spaces are similar, as shown by the correlation between weights vectors, with some inter-model variability.}
    \label{supp:fig:WeightsSOCAT}
\end{figure}

To assess the sensitivity of the learned feature importance to the sparsity of observations, we compared the feature weights obtained by optimizing the DII in two different spaces: the complete GOBM output, where samples are uniformly distributed in space and time, and the observational sampling defined by the SOCAT mask. Figure \ref{supp:fig:WeightsSOCAT} shows that the resulting feature weights are highly consistent across the two sampling strategies for all GOBMs considered, using $n = 13$ and $n = 7$, the minimal fully informative space, as case studies.  High correlation between the corresponding weight vectors indicates that the learned representations are largely insensitive to the sampling distribution, at least for these two domains.

\section{Inter-model comparison of GOBMs in uniform space.}
This section summarizes some information gained from studying the GOBMs in the full space where points are distributed uniformly in space and time. Some considerations apply to all models. 

1. The residual DII, shown as the purple bars in the right panel of Fig. \ref{fig:Domains}, is relatively small for all models, varying from $\sim$ 0.23 for the FESOM model to $\sim$ 0.27 for IPSL. This means that these input variables are good predictors of the chosen target. 

2. There is a plateau after $n$ = 7 in the DII curve (not shown), indicating that a near-optimal model of $\Delta$CO$_2$ is possible with 7 features; this is in line with what was observed in the domain defined by the SOCAT mask.

In Table \ref{supp:tab:rankings}, we show the ranking of features progressively eliminated during the optimal DII search as a function of $n$, for each GOBM. 
The overall picture is fairly consistent across models. The most important variables for all models, both in the full space and in the lower-dimensionality representations, are a combination of sea surface temperature (SST), latitude (A), salinity or salinity anomaly, and longitude information (B,C). The only exception is the MRI model, where the chlorophyll tracer variable, Chl, raises to the fourth position in ranking of importance. The variables xCO$_2$, Chl$_{an}$, T$_0$, and T$_1$ are consistently least important, and in fact essentially redundant, for all GOBMs. 

\begin{table}
\footnotesize
\setlength{\tabcolsep}{4pt}
\begin{tabular}{l|ccccccccccc}
\hline
 GOBM    &  1, 2, 3 & 4 & 5 & 6 & 7 & 8 & 9 & 10 & 11 & 12 & 13\\
 \hline
 ETHZ  & SST, A, C & SSS$_{an}$ & SSS & B & MLD & Chl & T$_0$ & T$_1$ & Chl$_{an}$ & SST$_{an}$ & xCO$_2$ \\ 
     \hline 
  FESOM & SST, A, SSS & C & B & SSS$_{an}$ & Chl & T$_0$ & MLD & T$_1$ & Chl$_{an}$ & xCO$_2$ & SST$_{an}$  \\
     \hline 
     NorESM & SST, SSS$_{an}$, A & C & SSS & B & Chl & T$_1$ & MLD & T$_1$ & SST$_{an}$ & Chl$_{an}$ & xCO$_2$ \\
     \hline 
     MRI & SST, A, C & Chl & B & SSS$_{an}$ & SSS & MLD & T$_0$ & T$_1$ & SST$_{an}$  & Chl$_{an}$ & xCO$_2$ \\ 
     \hline 
     IPSL & SST, A, B & C & SSS$_{an}$ & SSS & MLD & Chl & T$_0$ & T$_1$ & Chl$_{an}$ & SST$_{an}$ & xCO$_2$  \\
     \hline
\end{tabular}
\caption{Ranking of features progressively eliminated during the optimal DII search as a function of $n$, for each GOBM, using the full space where points are uniformly distributed in space and time. With some inter-model variations, the set of important and unimportant features is consistent across models.}
\label{supp:tab:rankings}
\end{table}

\begin{figure}
    \centering
    \includegraphics[width=\linewidth]{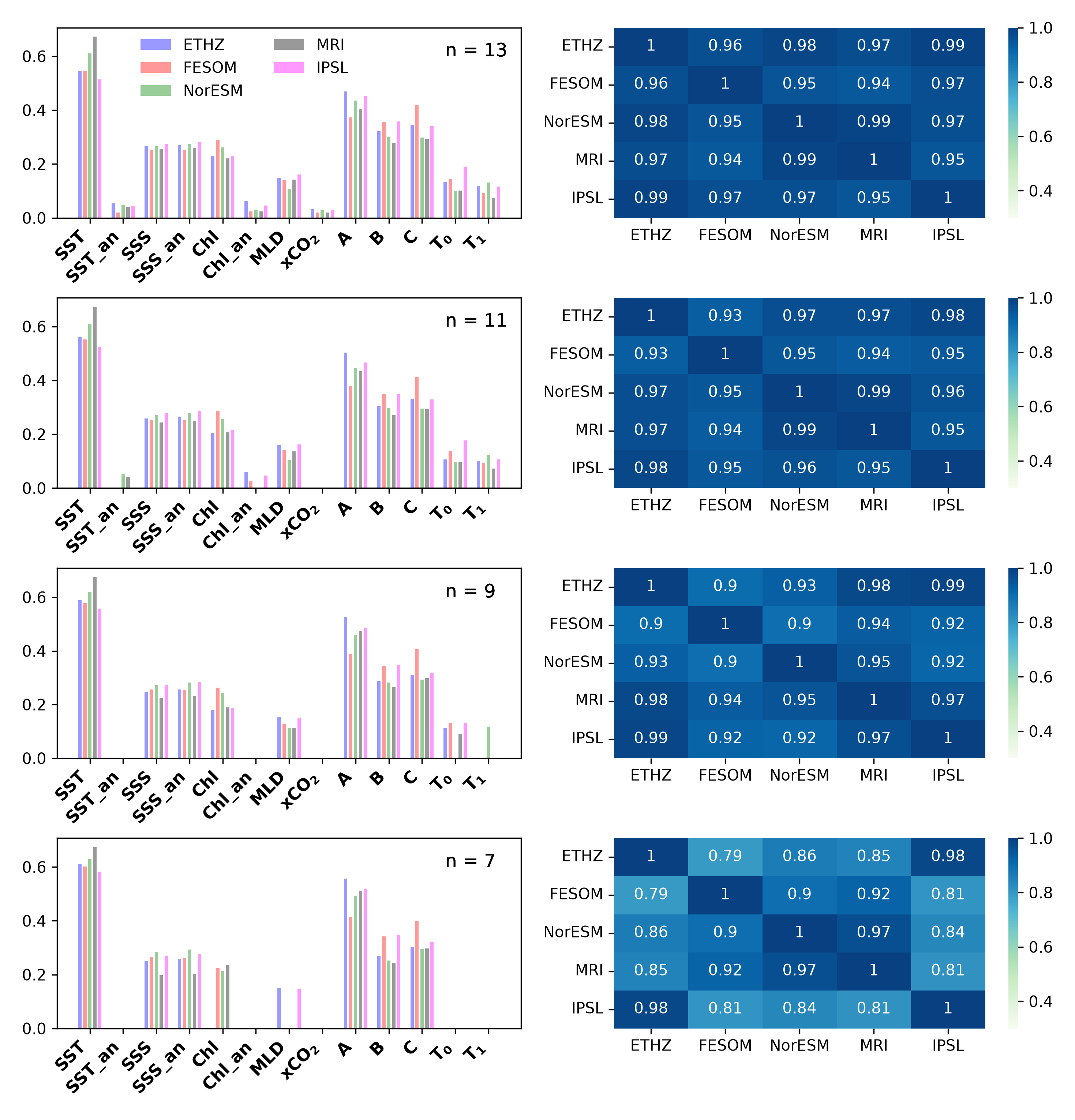}
    \caption{\textit{Left column:} Optimal weights found by the DII calculation for each feature and each GOBM, for $n$ = 13, 11, 9, and 7, using the full space where points are uniformly distributed in space and time. The sets of most important (SST, A, SSS/SSS$_{an}$, B, C) and least important (xCO$_2$, Chl$_{an}$, SST$_{an}$, T$_0$ and T$_1$) variables are consistent across models and dimensions. \textit{Right column:} Linear correlation coefficients between weights vectors at each $n$. These can be interpreted as a measure of similarity between GOBMs for each dimensionality of input space.}
    \label{supp:fig:Weights}
\end{figure}

Finally, in the left column of Figure \ref{supp:fig:Weights}, we show the (L2-normalized) weights found by the DII calculation for each feature and each GOBM, for progressively lower dimensionality of input space, from $n$ = 13 (the full space), to $n$ = 7 (the smallest near-optimal representation. In the right column of Fig. \ref{supp:fig:Weights}, we show the correlation coefficients between (non-zero pairs of) weights across different models. As discussed in the paper, this correlation can be interpreted as a metric of similarity. Overall, weights are fairly consistent across the landscape of models. Correlations among weights are high ($\rho > 0.94$) for all model pairs in the full space, but more inter-model differences emerge for smaller spaces (e.g., between ETHZ and FESOM and MRI and IPSL, where $\rho \sim 0.8$ for $n = 7$). 

\section{Re-weighting procedure for distribution matching}

\label{app:reweighting}

For the first generalization test described in Sec. \ref{sec:generalization_obs}, we wanted to create a near-uniform task to mimic the real-world task of CO$_2$ reconstruction in the full uniform domain.

Therefore, we sought a set of weights $\mathbf{w}$ for the source samples such that their weighted distribution matched a target distribution. The reweighting is performed in two stages.

First, we estimated an initial set of weights using unbalanced optimal transport (UOT) \cite{chizat2018scaling}. To reduce computational cost, the transport problem was solved on random subsets of the source and target samples. UOT computes a transport plan that minimizes the distance between the two distributions while allowing differences in total probability mass, making it well suited for cases in which the source and target distributions do not fully overlap. The resulting weights were then interpolated to the full source dataset using distance-weighted $k$-nearest-neighbor regression, clipped to remove extreme values, and normalized.

Second, the weights were refined using Iterative Proportional Fitting (IPF), also known as raking \cite{deming1940least}. IPF adjusts the weights so that the weighted source distribution better matches the target distribution along each individual feature dimension. During each iteration, the weights are updated according to the ratio between the target and weighted source histograms for each feature. Scaling factors are bounded to avoid excessively large weights, and the weights are clipped and renormalized after each iteration.

\begin{figure}
    \centering
    \includegraphics[width=\linewidth]{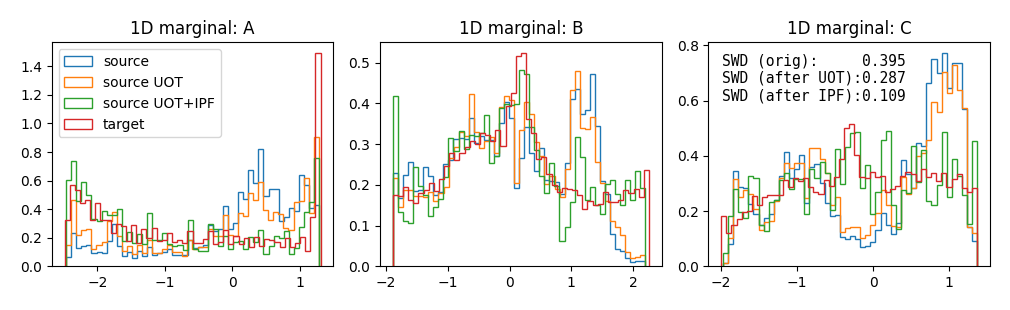}
    \caption{The two-stage re-weighting procedure described in the text allow us to match the 1D marginal distributions of source and target, allowing us to achieve a final value of SWD of 0.109, from an initial one of 0.395.}
    \label{fig:marginals}
\end{figure}

The quality of the reweighting was evaluated using the Sliced Wasserstein Distance (SWD) between the target samples and a weighted bootstrap sample drawn from the source distribution. We reported the SWD before reweighting, after the UOT initialization, and after the IPF refinement. We show the marginals for variables A, B, and C after each stage in Fig. \ref{fig:marginals}. The SWD calculated on the 3-dimensional distribution, also shown in the figure, decreases from an initial original value of 0.395 to a final value of 0.109.

\section*{Open Research Statement}

All the code used the generate data files and figures, as well as the intermediate data products they use, are made available at \\
\noindent
\url{https://github.com/vacquaviva/CO2RepresentCode}. Raw data files are large and we will share them upon request.

\section*{Conflict of Interest declaration}
The authors declare there are no conflicts of interest for this manuscript.


\section*{Acknowledgements}

VA acknowledges support from a PIVOT Research award (Award \#12871) from the Simons Foundation. VA, GAM, AF and THH acknowledge support from NSF through the Learning the Earth with Artificial intelligence and Physics (LEAP) Science and Technology Center (STC) (Award \#2019625) and from NOAA (Award \#NA24OARX431G0151-T1-01). VA thanks Gabriele Accarino and Davide Donno for sharing expertise on parallelization of sampling algorithms.



\bibliography{citations,Refs_IIpaper_2026}

\end{document}